\documentclass[reprint,amsmath,amssymb,aps,longbibliography]{revtex4-1}

\usepackage{graphicx}
\usepackage{bm}
\usepackage{hyperref}
\usepackage{amsmath}
\usepackage{braket}

\begin{document}

	\title{High-harmonic generation in Su-Schrieffer-Heeger chains}

	\author{Christoph J\"ur\ss}
	\affiliation{Institute of Physics, University of Rostock, 18051 Rostock, Germany }
	\author{Dieter Bauer}
	\affiliation{Institute of Physics, University of Rostock, 18051 Rostock, Germany }
	\date{\today}
	
	\begin{abstract}
          Su-Schrieffer-Heeger (SSH) chains are the simplest model
          systems that display topological edge states. We calculate
          high-harmonic spectra of SSH chains that are coupled to an
          external laser field of a frequency much smaller than the
          band gap. We find huge differences between the harmonic
          yield for the two topological phases, similar to recent
          results obtained with more demanding time-dependent density
          functional calculations [D.~Bauer, K.K.~Hansen, Phys.\ Rev.\
          Lett.\ {\bf 120}, 177401 (2018)]. This shows that the
          tight-binding  SSH model captures the essential
          topological aspects of the laser-chain interaction (while
          higher harmonics involving higher bands or screening in the
          metal phase are absent).  We study the robustness of the
          topological difference with respect to disorder, a
          continuous phase transition in position space, and on-site
          potentials. Further, we address the question whether the
          edges need to be illuminated by the laser for the huge
          difference in the harmonic spectra to be present.
	\end{abstract}

	\maketitle

	\section{Introduction} High-harmonic spectroscopy of condensed
        matter is an emergent field in strong-field attosecond
        science, which allows the all-optical probing of structural
        and dynamical properties \cite{Ghimire2011,SchubertO.2014,VampaPhysRevLett.115.193603,Hohenleutner2015,Luu2015,ndabashimiye_solid-state_2016,LangerF.2017,TancPhysRevLett.118.087403,you_high-harmonic_2017,Zhang2018,Vampa2018,Baudisch2018,Garg2018}. Topological phases became the
        focus of research directions such as topological insulators \cite{topinsRevModPhys.82.3045,topins,topinsshortcourse}, topological superconductivity \cite{TopolinsandsupconRevModPhys.83.1057,Viyuela2018}, cold atoms \cite{Atala2013}, topological photonics \cite{Rechtsman2013,Stutzer2018,Kruk2019}, toplogical electronic circuitry \cite{PhysRevX.5.021031,PhysRevLett.114.173902,Wang2019}, and topological mechanics \cite{Kane2013}. Only very recently, the exploration of the physics at the interface between strong-field attosecond science and topological condensed matter began theoretically \cite{PhysRevB.96.075409,bauer_high-harmonic_2018,silva_all_2018,chacon_observing_2018,2019arXiv190101437D,PhysRevA.98.063426} and experimentally \cite{Luu2018,Reimann2018}.   Of particular interest there is the all-optical distinction of topological phases, the steering of electrons through Berry curvatures or along topologically protected edges on sub-laser-cycle time scales, with potential applications in coherent light-wave electronics \cite{Sommer2016,Garg2016,Higuchi2017,PhysRevLett.121.207401}.

\begin{figure} 
		\includegraphics[width=0.8\columnwidth]{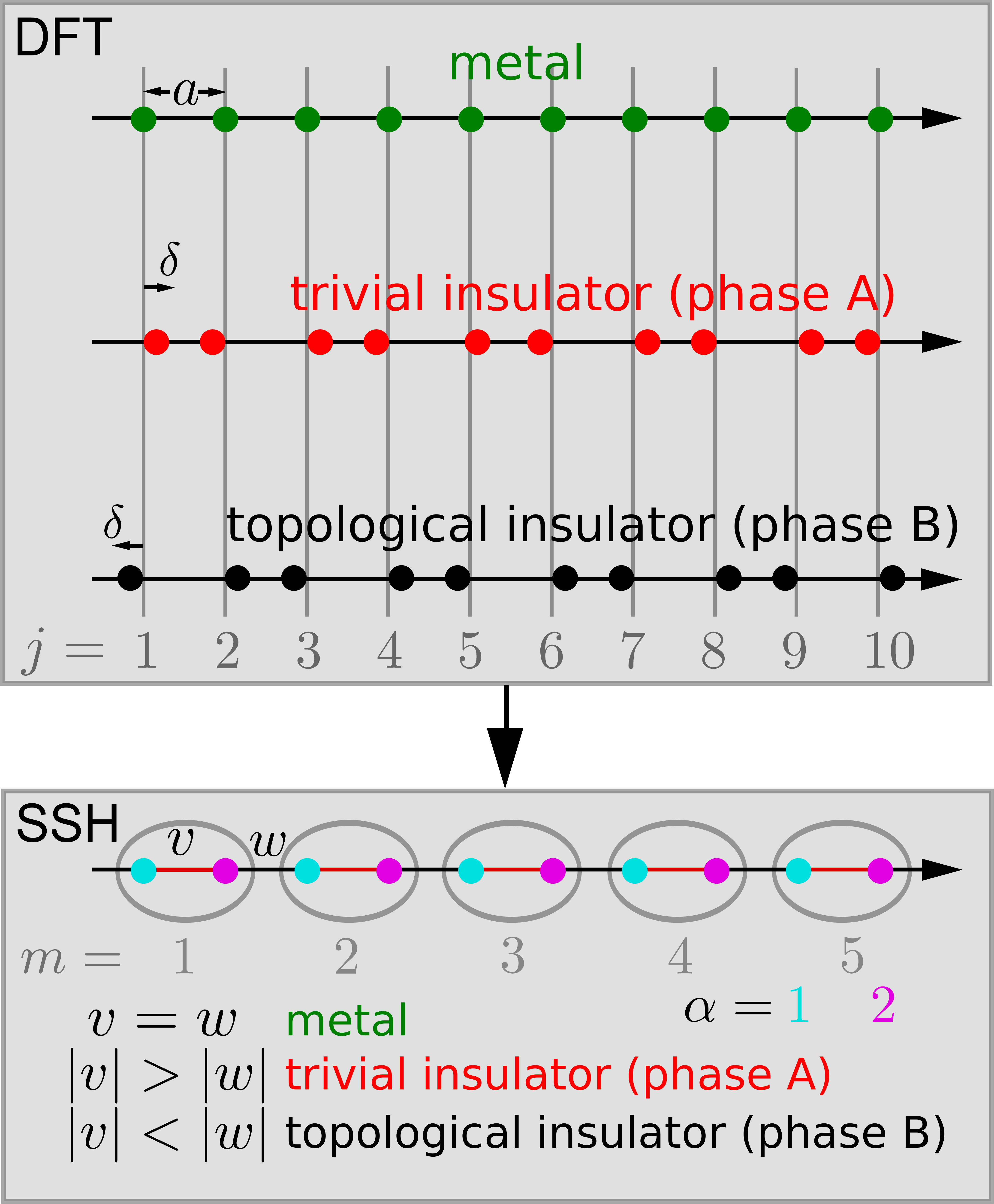}
		
\caption{\label{fig1} Connection between atom positions on a fine-grained, real-space grid as used in DFT \cite{PhysRevA.96.053418,bauer_high-harmonic_2018,PhysRevA.97.043424,PhysRevA.99.013435,2019arXiv190101437D} and the SSH
model \cite{SSHPhysRevLett.42.1698,PSSB:PSSB2221270102,topinsshortcourse} to illustrate the discussion in the Introduction.  }
	\end{figure}

In Ref.~\cite{bauer_high-harmonic_2018}, harmonic generation in
dimerizing linear chains was investigated using time-dependent density
functional theory (TDDFT) \cite{rungegross84,ullrich_time-dependent_2011}. A huge difference in the harmonic yield for
the topological phases A and B (see Fig.~\ref{fig1}) was observed and
attributed to the presence of topological edge states in phase B.  The
band structures resembled qualitatively those known for the
Su-Schrieffer-Heeger (SSH) model \cite{SSHPhysRevLett.42.1698}, originally introduced as a tight-binding model for polyacetylene \cite{PSSB:PSSB2221270102,BLOCK199631} (see
Ref.~\cite{topinsshortcourse} for a modern introduction into the
topological aspects of the SSH model).
	
Figure~\ref{fig1} illustrates the connection between the modelling
using the very simple SSH  tight-binding approach (leading to two bands only) \cite{SSHPhysRevLett.42.1698,PSSB:PSSB2221270102,topinsshortcourse,gebhardPhMB} and the ab-initio density-functional theory (DFT) on a fine-grained position-space grid \cite{PhysRevA.96.053418,bauer_high-harmonic_2018,PhysRevA.97.043424,PhysRevA.99.013435,2019arXiv190101437D}. Let us first consider the upper panel. The atoms are
shifted from their equidistant positions (lattice constant $a$)
alternatingly by $\delta$ to the right and left, generating the two
possible dimerizations called phase A and phase B.  For the case of
one electron per ion, the equidistant configuration is metallic (half
populated lowest band) but energetically less favorable than the
dimerized phases (Peierls instability). A band gap
opens  for phases A and B (metal-to-insulator Peierls transition) because the lattice
constant doubles, i.e., the Brillouin zone halves, and the half
populated lowest band of the metal becomes a fully populated valence
band. Phase B has (for an even number of ions $N$ in the chain) two
edge-ions without partner ion to dimerize with. This leads to
topological edge states in the band gap between valence and
conduction band (see Fig.~1(b-d) in Ref.~\cite{bauer_high-harmonic_2018}
for the DFT model).

The lower part of Fig.~\ref{fig1} illustrates the SSH
model. The electronic part of the free SSH Hamiltonian  [see
Eq.~(\ref{H_SSH}) or  \eqref{eq:HSSHapp}] allows for intra-cell hopping with
amplitude $v$ between the two lattice sites $\alpha=1,2$
within a primitive cell $m=1,2, \ldots N/2$, and for inter-cell hopping
with amplitude $w$. A connection between the DFT and the SSH
model can be established in the spirit of a tight-binding
approximation. Both phase A and phase B display two internuclear
distances $a-2\delta$ and $a+2\delta$ in the DFT model. As a
consequence, the tunneling of electrons between neighboring atoms is more
likely for the smaller distance and less likely for the larger. In the
SSH model, this is taken into account by the two hopping elements $v$
and $w$. The three cases $v=w$, $|v|>|w|$, and $|w|>|v|$ correspond to
metal, phase A, and phase B, respectively.  Note that the SSH model
describes a bi-partite system because only hopping between sites with
different $\alpha$ are allowed.

	The SSH model is much simpler than the DFT model. In fact, the
SSH model assumes non-interacting electrons and hence reduces to a
single-electron problem with only nearest-neighbor hoppings and absent
on-site interaction. Because of this simplicity, all the essential
features of the SSH model (e.g., band structure, winding number, edge states) can
be derived analytically (see, e.g., \cite{topinsshortcourse}). 

A short introduction to the SSH model and its coupling to an external field are
given in section~\ref{sec:theory}.   In section~\ref{sec:results}, first the harmonic
spectra for unperturbed SSH chains are discussed before, in section~\ref{sec:randomshifts}--\ref{sec:nonvanishingonsite}, the robustness of the huge difference in the harmonic yield due to topological edge states is investigated with respect to random shifts of the atoms in the chain, a continuous transition  between phase A and B in position space, and non-vanishing on-site potential, respectively. Finally, spectra for a hypothetically localized laser field are presented in section~\ref{sec:BBC} in order to address the question whether the laser needs to illuminate the edges to reveal the huge difference between the two topological phases.

	\section{Theory}\label{sec:theory}
		
		\subsection{The Su-Schrieffer-Heeger model}
The SSH tight-binding Hamiltonian belongs to the wider class of models for dimerized quantum chains \cite{PhysRevB.97.144412}. It was originally introduced to describe polyacetylene using a tight-binding description for the $p_z$ electrons and elastically coupled CH monomers  \cite{SSHPhysRevLett.42.1698,PSSB:PSSB2221270102}. We are not interested in the ion dynamics (i.e., phonons) and just consider the electronic part of the SSH Hamiltonian for a given ion configuration. Moreover, we are not thinking of polyacetylene specifically but any 1D chain (or 2 or 3D systems along the laser polarization direction) and thus write $N$ ``sites'' (instead of $N$ CH monomers).
For an even number of sites $N$, the SSH model
consists of $n = N/2$ primitive cells with two lattice sites
$\alpha = 1,2$ each. The real-valued hopping elements $v$, $w$
describe the intra-cell and inter-cell hopping of an electron, respectively. Without
external field, the electronic SSH Hamiltonian matrix reads
			\begin{equation}\label{H_SSH} \mathbf{H}_0
= \begin{pmatrix} 0 & v & &&&&& \\ v & 0 & w &&&& \\ & w&0&v&&&& \\
&&v&0&w&&& \\ &&&\ddots&\ddots&\ddots& \\ &&&& w& 0&v \\ &&&&&v&0
				 \end{pmatrix}.
			\end{equation}
	This Hamiltonian has $N$ eigenstates
			\begin{equation}
\Psi_i=\left(\Psi_i^1,\Psi_i^2,...,\Psi_i^j,...,\Psi_i^N\right)^\top,
			\end{equation} with $i = 0,1,...,N-1$ where
$\Psi_i^j$ is the value of the electronic wavefunction at site $j=1,2, \ldots, N$. The electron is on a lattice site with $\alpha = 1$ ($2$) if $j$ is odd (even).  The
Hamiltonian \eqref{H_SSH}  has chiral symmetry \cite{topinsshortcourse} and thus a
symmetric energy spectrum, i.e.,  the energies of the eigenstates (in ascending order) fulfill 
 $E_i = -E_{N-1-i}$ for $i=0,1,...,N/2-1$.

For  periodic
boundary conditions (i.e., bulk or rings), the electronic SSH Hamiltonian matrix
reads
\begin{equation}\label{H_SSH_bulk}
\mathbf{H}_0^{(\mathrm{bulk})} = \begin{pmatrix} 0 & v & &&w \\ v & 0
& w && \\ &\ddots&\ddots&\ddots& \\ && w& 0&v \\ w&&&v&0
			\end{pmatrix}.
			\end{equation}
			Eigenstates for the bulk system can be
derived analytically (see appendix~\ref{app:A}). We choose $v,w < 0$ so that the  number of nodes in the energy eigenstates  increases with energy.

		\subsection{Position information and coupling to
external field}
		
Accepting the hopping elements $v$ and $w$ as free parameters,  the electronic SSH model does not require any information
about the position of the atoms. However, this information is needed
for the coupling to an external field. We make the same choice of the
atomic-site positions $x_j$ as in \cite{bauer_high-harmonic_2018}   [atomic units (a.u.) $\hbar=|e|=m_e=4\pi\epsilon_0=1$ are used]:		
			\begin{equation} x_j = \left(j -
\frac{N+1}{2}\right) a - (-1)^j\delta, \quad j = 1,2,...,N.
			\end{equation}
			Here,  $j$ is the atomic-site index, $a$ is the distance
between the atoms in the metallic case $\delta=0$, and $\delta\neq 0$ describes the alternating shift of the atoms causing the dimerization (see Fig.~\ref{fig1}). We assume that the tunneling probability between neighboring sites scales
exponentially with distance and set the hopping elements to			
			\begin{equation}\label{v} v =
-\exp[-(x_2-x_1)]=-\exp[-(a-2\delta)],
			\end{equation}
			\begin{equation}\label{w} w =-\exp[-(x_3-x_2)]
=-\exp[-(a+2\delta)].
			\end{equation}
			
\begin{figure}
				\includegraphics[width=1\columnwidth]{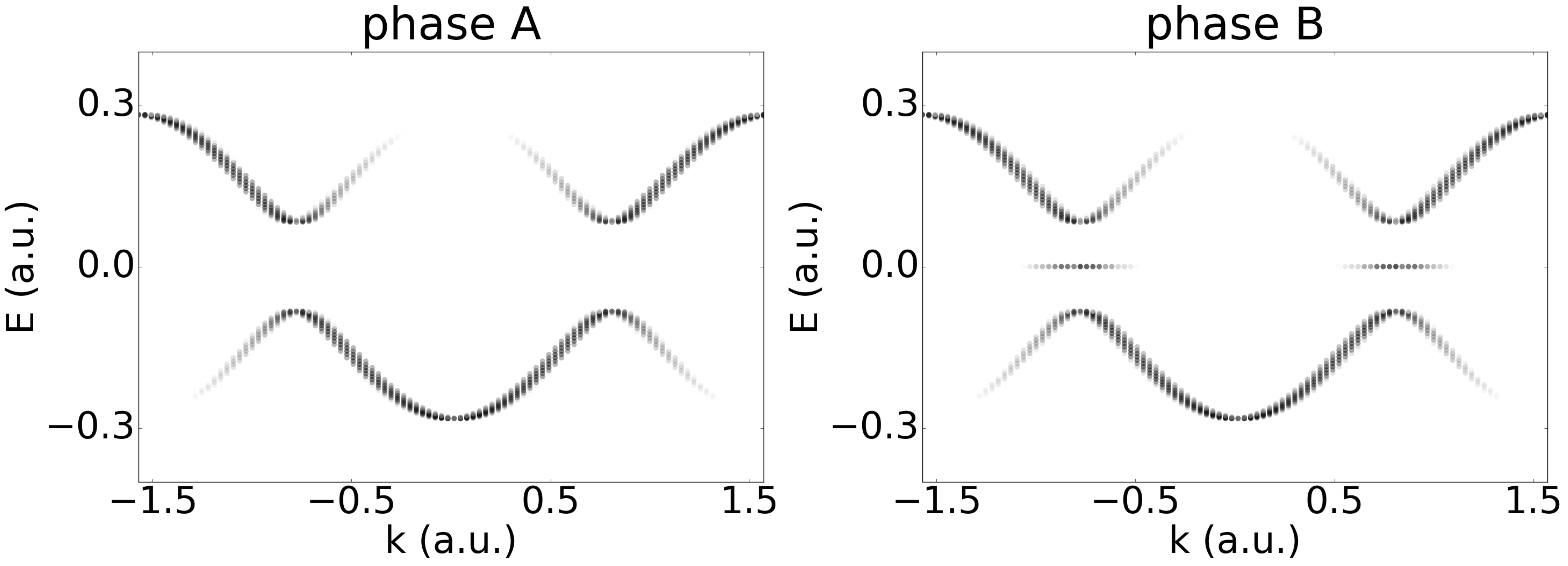}
				
				\caption{Band structures for $N=100$, $a=2$, and $\delta=0.15$ (trivial-insulator phase A, left) and $\delta=-0.15$ (topological-insulator phase B, right). The first Brillouin zone for the metallic case $\delta=0$ would be $[-\pi/a,\pi/a]$. For the energetically favorable, dimerized case $\delta\neq 0$, the Brillouin zone halves (Peierls instability). }
				\label{fig:bands_AB}
			\end{figure}

  For $\delta>0$  ($\delta<0$), the system is in phase
A (B).  As described in more detail in
Ref.~\cite{topinsshortcourse}, the system consists of two bands if
$\delta \neq 0$, see Fig.~\ref{fig:bands_AB} \footnote{The ``band structure'' for finite SSH chains is calculated by Fourier-transforming the eigenstates, i.e., $\Psi_i(x)\rightarrow\tilde\Psi_i(k)$, and plotting $\log|\tilde\Psi_i(k)|^2$ vs $k$ and the respective energy $E_i$ as color-coded contours. The step size for $k$ is $\Delta k = \frac{2\pi}{N a}$. Hence possible $k$-values are $k = 0,\Delta k, 2\Delta k,...,(N-1)\Delta k$. The first Brillouin-zone in the metal case is $[-\frac{\pi}{a},\frac{\pi}{a}]$. In phases A and B the spacing between the atoms is not equidistant. However, as long as $|\delta| \ll a$ we can use the same procedure to calculate the band structures shown in Fig.~\ref{fig:bands_AB} for illustration.}.

For one electron per site, the  lower band (valence band) is fully populated while the upper band (conduction
band) is empty.  The band gap between them increases with the absolute value of
$\delta$, independent of the sign. But for $\delta<0$ (phase B) there are two additional states in the middle of the band gap. These almost degenerate states around zero energy are spatially localized at the edges of the chain (hence edge states), one of them being odd, the other even with respect to inversion about the origin $x=0$. In the limit $N\to\infty$ the states become exactly degenerate zero-energy states \footnote{Due to the lack of chiral symmetry of the Kohn-Sham Hamiltonian, the conduction and valence band are neither symmetric about energy $E=0$ nor are the edge states in phase B exactly in the middle of the band gap in the DFT band structure in Ref.~\cite{bauer_high-harmonic_2018}. This shows already that chiral symmetry and the related existence of a winding number \cite{topinsshortcourse} is not necessary for a 1D chain to display degenerate edge states.}.

			The chain is coupled to a linearly
polarized laser field in dipole approximation,		
			\begin{equation}\label{E_A} \mathbf{A}(t) =
A(t)\mathbf{e}_x,~~~~~\mathbf{E}(t) = -\partial_t\mathbf{A}(t).
			\end{equation}
The dipole approximation is adequate because we assume that the linear chain is parallel to the laser polarization direction and small compared to the focus of the laser pulse. 
		  In Ref.~\cite{PhysRevB.51.4940},
a gauge-invariant coupling of external drivers to tight-binding
models was presented. In length gauge, the diagonal elements $H_{jj}$ need to be replaced
according			
			\begin{equation} H_{jj} = 0 \longrightarrow
-\Phi(x_j,t) = E(t) x_j,
			\end{equation}			
		whereas in velocity gauge, the hopping elements become
			\begin{equation} H_{jl}\longrightarrow
H_{jl}\, e^{-i(x_j-x_l) A(t)}.
			\end{equation}						
			In length-gauge, the Hamiltonian matrix thus reads
			\begin{equation} \mathbf{H_{LG}}(t)
= \begin{pmatrix} E(t)x_1 & v & && \\ v & E(t)x_2 & w && \\
&w&E(t)x_3&v& \\ &&&\ddots& \\ &&&v&E(t)x_n
			\end{pmatrix}.
			\end{equation}
For periodic boundary conditions, this implies a
discontinuous scalar potential because $E(t)x_{N+1}\neq E(t)x_1$. In
that case one has to use velocity gauge,			
			\begin{equation} \mathbf{H_{VG}}(t)
= \begin{pmatrix} 0 & v^*(t) & && w(t)\\ v(t) & 0 & w^*(t) && \\
&w(t)&0&v^*(t)& \\ &&&\ddots& \\  w^*(t)&&&v(t)&0
			\end{pmatrix}, \label{eq:HVG}
			\end{equation}
			with
			\begin{align} v(t) &= v
\exp[{-i(a-2\delta)A(t)}] \nonumber \\ &= -\exp\{-(a-2\delta)[1+iA(t)]\}, \\
 w(t) &= w \exp[{-i(a+2\delta)A(t)}] \nonumber \\ & = -\exp\{-(a+2\delta)[1+iA(t)]\}. 
			\end{align}
One may also use velocity gauge for finite chains without periodic boundary; in that case the upper right and lower left corner elements in \eqref{eq:HVG} are absent.	
			The gauge-invariant coupling in velocity gauge proposed in
Ref.~\cite{PhysRevB.51.4940} reduces to the usual Peierls
substitution in our case with dipole approximation (see appendix~\ref{app:gauge}).

	\subsection{Numerical calculations}
	The eigenstates of the $N$-dimensional SSH Hamiltonian
(\ref{H_SSH}) are obtained by diagonalization. The $N/2$
lowest energy states (occupied by $N$ electrons, assuming spin degeneracy) are propagated in time from
the beginning to the end of a laser pulse.  An $n_\mathrm{cyc} = 5$-cycle sine-squared
laser pulse is used with
		\begin{equation} A(t) = A_0
\sin^2\left(\frac{\omega t}{2 n_\mathrm{cyc}}\right)
\sin\omega t,~~~0<t<2\pi\omega/n_\mathrm{cyc} \label{eq:pulse}
		\end{equation}		
		and zero otherwise. The electric field follows from \eqref{E_A}. 
The frequency is set to $\omega = 0.0075$ (i.e., $\lambda\simeq 6.1\,\mu$m), and
the vector potential amplitude is $A_0 = 0.2$ (corresponding to a laser intensity of $\simeq 7.9\times 10^{10}$\,Wcm$^{-2}$) throughout the paper. The results discussed in this paper are qualitatively insensitive to the details of the laser pulse as long as $A_0$ is large enough to yield high harmonics at all, and the laser frequency is small compared to the band gap. 
		
		Wavefunctions are propagated in time using the
Crank-Nicolson approximant to the time-evolution operator
		\begin{equation} \exp[-i\mathbf{H}(t)\Delta t]
 = \frac{1-i\mathbf{H}(t)\Delta t
/2}{1+i\mathbf{H}(t)\Delta t /2} + {\cal O}(\Delta t^3),
		\end{equation}		
		where the discrete time step is set to $\Delta t =
0.1$.		High-harmonic spectra for the finite chains may be calculated from the dipole, the acceleration, or the current \cite{PhysRevA.79.023403,0953-4075-44-11-115601}, differing by prefactors $\omega^4$ or $\omega^2$, respectively.  Apart from the sign, the position expectation value equals the dipole and reads
		\begin{equation} X(t) = \sum_{i = 0}^{N/2 -
1}\sum_{j=1}^{N} \Psi_i^{j*}(t) x_j \Psi_i^j(t)
		\end{equation}	
		where $i$ labels the state and $j$ the
position. Semi-classically and for uncorrelated emitters, the spectrum of the radiated light $P(\omega)$ is
proportional to	the absolute square of the Fourier-transformed dipole acceleration \cite{PhysRevA.41.6571}, 	
		\begin{equation} P(\omega) \propto \left| \mathrm{FFT}\left[
\ddot{X}(t)\right]\right|^2.
		\end{equation}
	We normalize the spectra to the maximum of 
		
		\begin{equation}
		P_{\mathrm{free}}(\omega) = \left| \mathrm{FFT}\left[
		E(t)\right]\right|^2 .  \label{eq:Pfree}
		\end{equation}
	
	Since for bulk calculations with periodic boundary conditions the dipole is not defined and the length gauge cannot be used, harmonic spectra were determined from the current $I(t)$ in velocity gauge as $	P(\omega) \propto \left| \mathrm{FFT}\left[I(t)\right]\right|^2$.	
	However, there is no difference between harmonic spectra $P(\omega)$ from phase A and phase B for periodic boundary conditions. On one hand, one may expect this because phase A and B look alike for periodic boundary conditions. On the other hand,  one may expect a difference because of the bulk-boundary correspondence \cite{topinsRevModPhys.82.3045,PhysRevB.97.115143}. We address this issue in section~\ref{sec:BBC}.

	\section{Results}\label{sec:results}
\begin{figure} \centering
				\includegraphics[width=0.8\columnwidth]{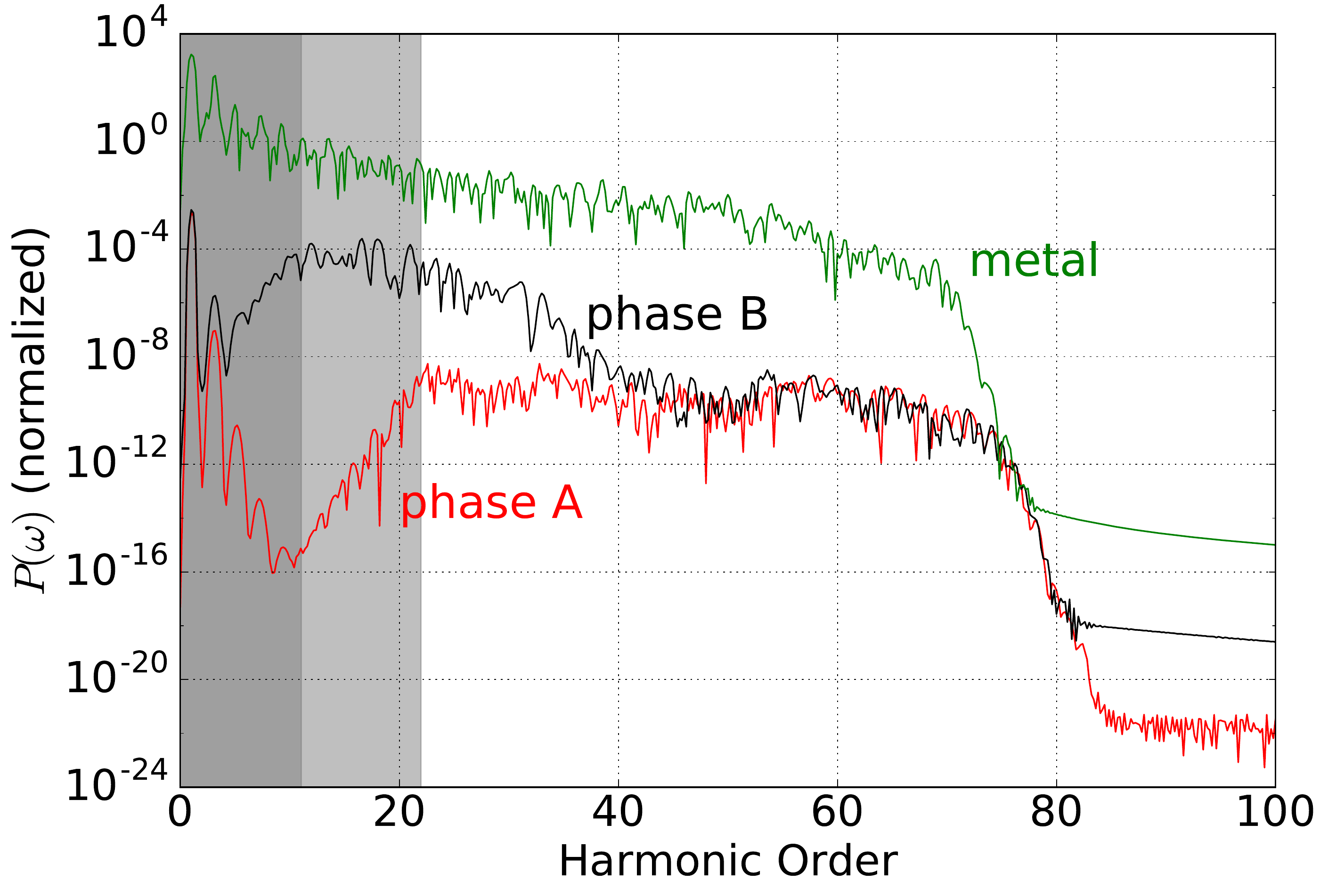}
				\caption{Harmonic spectra for the two
topological phases A and B, and the metallic case. The lighter gray-shaded area
indicates harmonic photon energies smaller than the band gap for phase
A, the darker gray-shaded area indicates energies below the gap between bands and
edge states in phase B. The harmonic yield $P(\omega)$ is normalized to $P_{\mathrm{free}}(\omega)$, eq.~\eqref{eq:Pfree}. }
				\label{fig:all_phases}
			\end{figure}

 Consider a chain with $N=100$ sites, $a=2$, and $\delta=0.15$ (phase A), $\delta=-0.15$ (phase B). We will focus on this specific chain configuration throughout the paper if not stated otherwise. For the sake of completeness, we also consider the metallic case $\delta=0$.  The corresponding harmonic spectra for the laser pulse  \eqref{eq:pulse} are presented in Fig.~\ref{fig:all_phases}. Similar to the results obtained with TDDFT \cite{bauer_high-harmonic_2018}, we find a huge
difference in the harmonic yield for phases A and B for harmonic orders smaller than the
band gap $\Delta E_{gap}$, corresponding to  harmonic order $\simeq 22$. Harmonics above the band gap are so-called inter-band harmonics and  produced in the usual three-step way known from the gas phase \cite{vampa_merge_2017}: an electron tunnels into to the conduction band, electron and hole move in the conduction and valence band, respectively, and recombine when they meet in position space, upon emission of harmonic radiation.  The ultimate highest harmonic in a tight-binding system such as the SSH model thus is the maximum energy difference between valence and conduction band, which is $\Delta E_{\max}\simeq 0.566$ for phases A and B, and $0.541$ for the metal.  In terms of harmonic orders this corresponds to $\Delta E_{\max}/\omega=75.4$ and $72.2$, respectively, which agrees with the ultimate cut-offs in Fig.~\ref{fig:all_phases}. If higher conduction bands are taken into account (as done in TDDFT simulations), harmonics beyond these cut-offs can be generated. 
Below the band gap (i.e., for harmonic orders $< 22$),  the harmonic yield for phase A drops because the efficient three-step mechanism cannot produce harmonics at such low energies. Only clean intra-band harmonics for orders $\leq 9$ are observed, which originate from the motion of electrons in the non-parabolic regions of the valence band. For phase B, we find high-harmonic yield down to harmonic order $11$, which corresponds to the energy difference
between the valence band and the edge states (half the
band gap of phase A). These low harmonics can be generated via electronic transitions between edge states and valence band. Below harmonic order $11$, the yield for phase B decreases a bit before intra-band harmonic generation takes over, as in phase A. 

Intra-band harmonics from a fully populated valence band tend to interfere away because in such a simple band structure  as the one for the SSH chain, for each valence-band electron initially located at a $k$-point with a certain band curvature there is another electron with the opposite band curvature. Such pairs of electrons oscillate with opposite excursions when driven by the laser field so that their dipole radiation interferes destructively. That is the reason why the intra-band harmonics drop rapidly with the harmonic order both for phase A and B in Fig.~\ref{fig:all_phases}. However, from harmonic order $11$ on, transitions to the edge states come into play for phase B, and the harmonic yield increases again. For phase A, inter-band harmonic generation only sets in towards band-gap harmonic order 22. As a consequence, a huge difference in the harmonic yield for harmonic orders $\in [\Delta E_{\mathrm{gap}}/2\omega,\Delta E_{\mathrm{gap}}/\omega]$ arise, explaining the observations obtained with TDDFT in Ref.~\cite{bauer_high-harmonic_2018} but with a simpler tight-binding model. The differences for energies above the band gap up to order $\approx 40~(\approx\Delta E_{max}/2\omega = 38)$ are also due to the edge states present in phase B. With edge states in the band gap, transitions to the conduction  band  are more likely, leading  to a higher harmonic yield for phase B.

			For completeness, the spectrum for the metallic
phase is plotted in Fig.~\ref{fig:all_phases} as well. The metal has no band gap and only a single, half-occupied band so that the cancellation due to opposite band curvature does not take place. As a result, one observes efficient harmonic generation up to the ultimate cut-off energy. However, because of the absence of screening in the SSH model this is not a realistic description of harmonic generation in a metal. In fact, the spectrum for the metal case obtained with TDDFT in \cite{bauer_high-harmonic_2018}, where screening is taken into account, does not show efficient harmonic generation \footnote{Whereas a TDDFT simulation with ``frozen'' (i.e., ground-state) Kohn-Sham potential does.}.

For periodic boundary conditions (i.e., bulk or a ring chain) the spectra for both
phases are identical and the dip below the band gap is comparable to
the one of phase A in a finite chain.

	The hopping elements as a function of the 
distance between neighboring atomic sites are defined in  eqs.~\eqref{v} and \eqref{w}. In the
following, we initialize the atomic-site positions in various ways that deviate from the pure, dimerized cases A and B in order to study the robustness of the band structure and the harmonic spectra for the respective configurations. 
	
		\subsection{Random shifts}		 \label{sec:randomshifts}
\begin{figure} \centering
				\includegraphics[width=1.0\columnwidth]{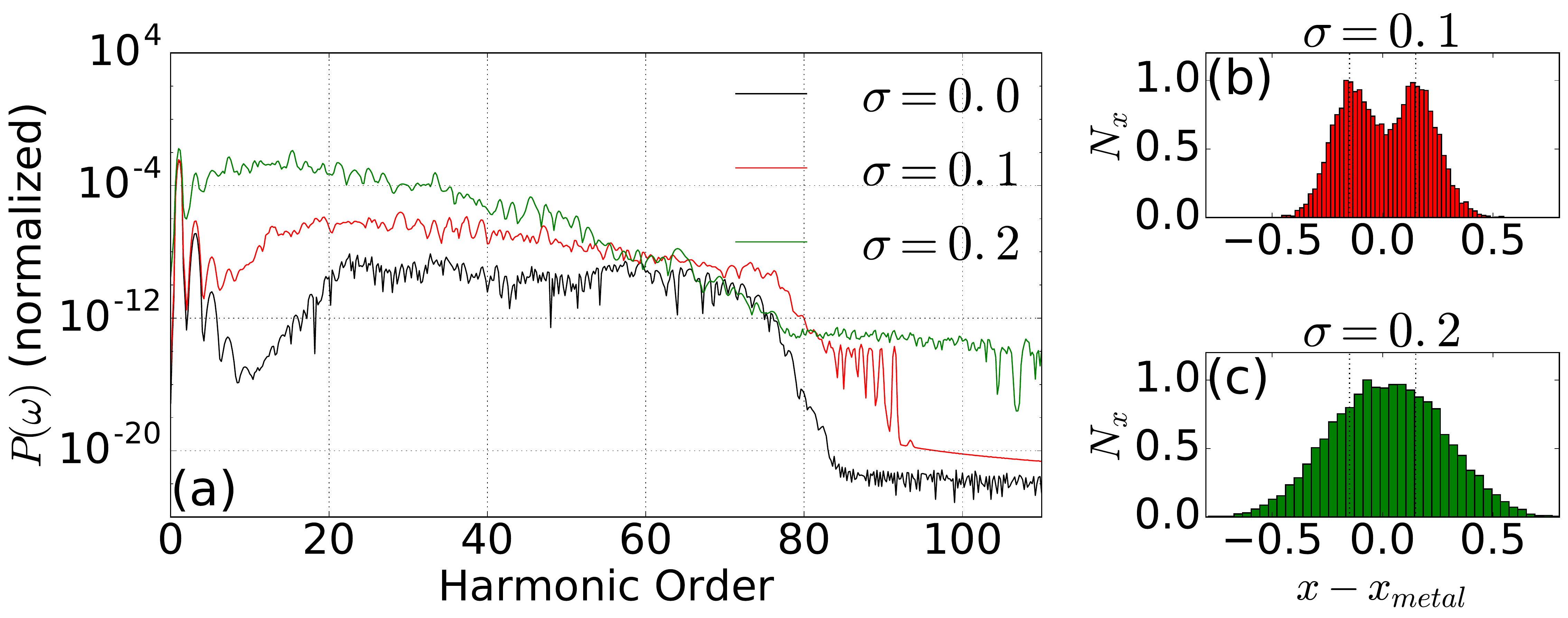}
				\caption{(a) Averaged harmonic spectra over 100 ensembles of randomly shifted atomic sites with respect to the unperturbed phase-A configuration for $\delta=0.15$. The shifts follow a normal distribution of variance $\sigma$. Panels (b) and (c) show the deviations of $x_i$ from the metallic case $x_{metal}$ as histograms. For the unperturbed phase A, all values would be located at $\pm\delta=\pm0.15$, indicated by vertical dashed lines.}
				\label{fig:rand_deltaA}
			\end{figure}

Starting from the pure phase-A case with $\delta=0.15$, we shift each atom from its original position $x_i^0$ by a random $\Delta x_i$ to $x_i = x_i^0 + \Delta x_i$ in order to investigate the influence of disorder on the harmonic spectra. These random shifts cause a modification of the hopping amplitudes. The rate to jump from atom $i$ to $i+1$ (and back) is given by 
\begin{equation}\label{hopping_gen}
	t_{i,i+1} = t_{i+1,i} = -\exp[-|x_{i+1}-x_i|].
\end{equation}
The random shifts obey a normal distribution of variance $\sigma$. We calculate averaged spectra for $\sigma = 0.1$ and $0.2$ with ensembles of 100 configurations each. The averaged spectra are shown in Fig.~\ref{fig:rand_deltaA} (a), together with the
unperturbed case $\sigma = 0$ for reference.
			
A decreased harmonic yield at low harmonic orders for a variance $\sigma = 0.1$ can be still
observed while the dip disappears for $\sigma = 0.2$. The deviations of the atomic positions from the metal case are shown in Fig.~\ref{fig:rand_deltaA} (b) and (c) as histograms. For the pure phase A the deviations are either $-\delta=-0.15$ or $+\delta = +0.15$. The distribution for  $\sigma = 0.1= 2\delta/3$ is already quite broadened but two maxima are still clearly visible. For $\sigma = 0.2$, the atom positions are too random to yield two maxima in histogram. 

The energies of the states explain some features of the spectra. With increasing $\sigma$,  the maximum energy difference in general increases, causing a higher ultimate cut-off in the spectrum. In addition, the band gap closes.  
The disappearance of the dip in
the spectrum for larger variances is therefore caused by the disappearing
band gap. 
However, this closing of the band gap due to disorder  happens surprisingly slowly as a function of increasing $\sigma$. For a variance $\sigma = 0.1= 2\delta/3$, the band gap is still
clearly visible, and  the dip in the harmonic spectrum of
phase A is thus remarkably robust against disorder in the atomic positions.
			   
The same happens for phase B (not shown) although the dip there is less pronounced in the first place because of the edge states that effectively halve the band gap.

		\subsection{Phase transition} \label{sec:phasetransition}	
\begin{figure} \centering
				\includegraphics[width=0.75\columnwidth]{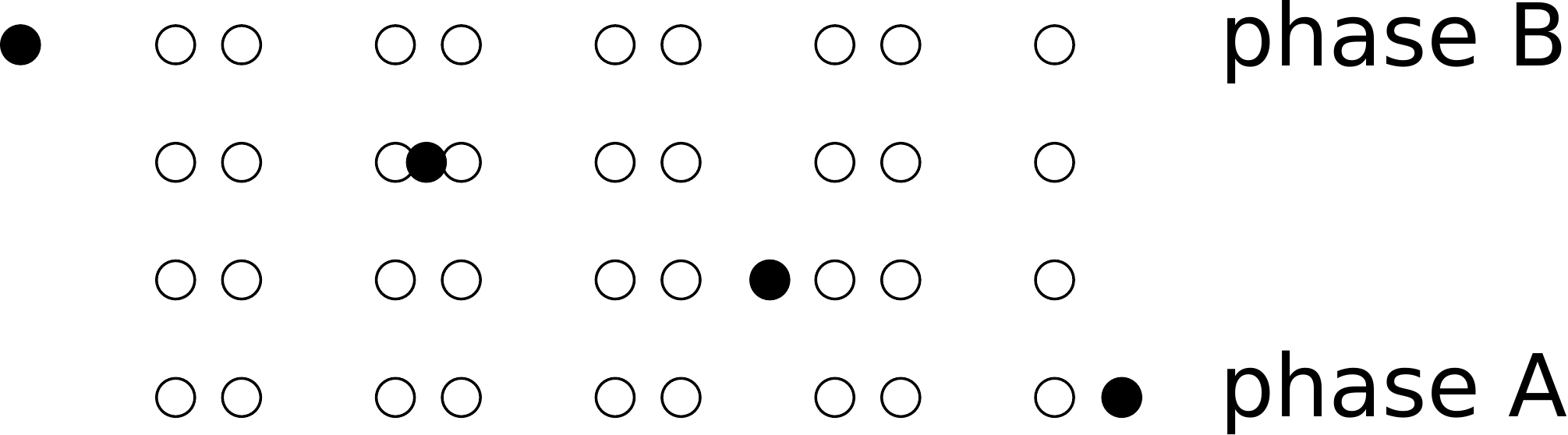}
				\caption{Illustration of the continuous phase
transition between the dimerized chain in phase B to phase A in position space (for better visibility only $N=10$ atoms are considered). The filled circle indicates the moving
atom whose position is $x_1$. The other circles indicate the fixed
atoms.}
				\label{fig:sketch}
			\end{figure}

\begin{figure} \centering
				\includegraphics[width=1.0\columnwidth]{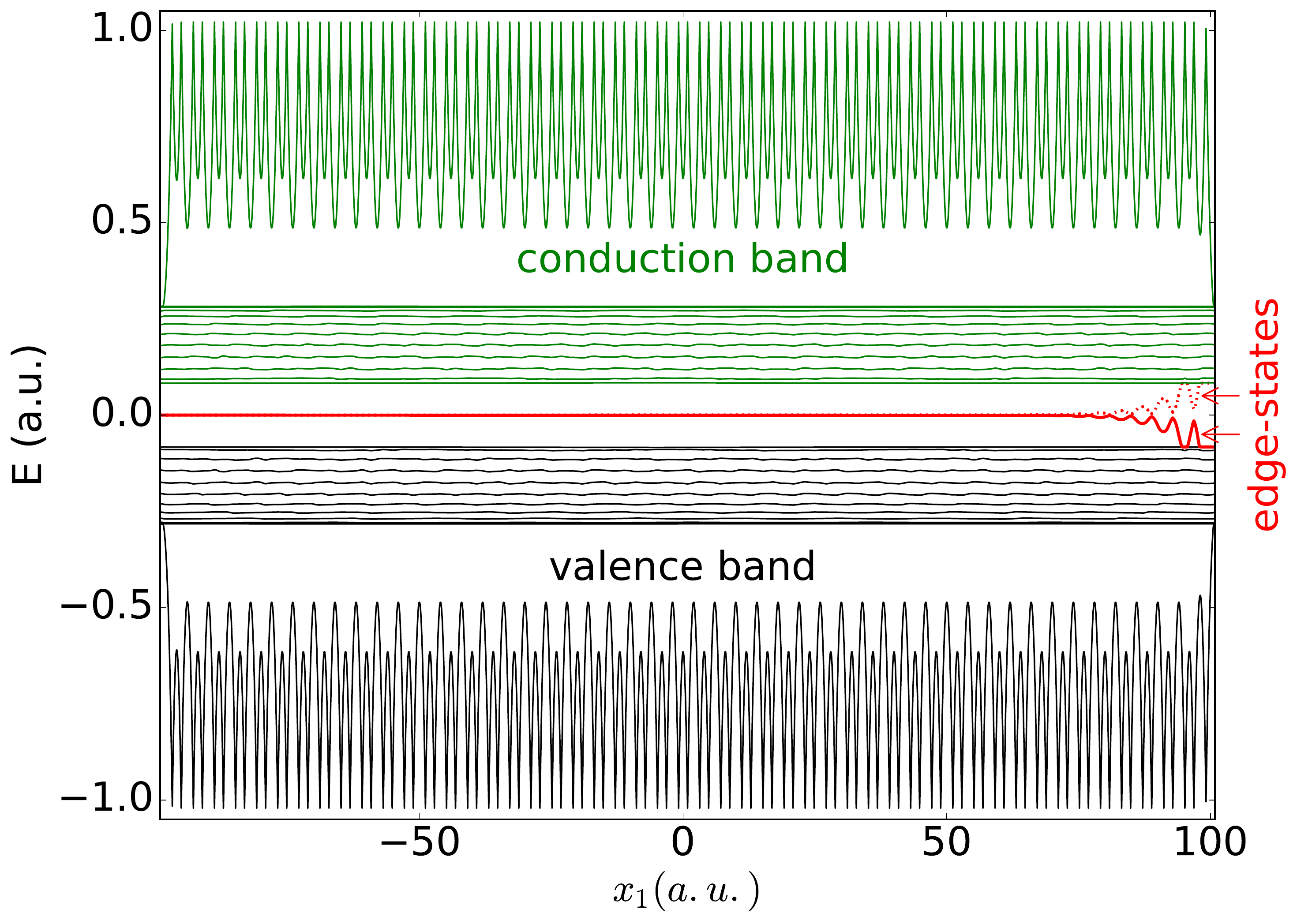}
				\caption{SSH eigenenergies as a function of the position $x_1$ of the shifted ion that is originally at the left edge. The energies of the two initially degenerate edge states depart only from $E=0$ when $x_1$ is already close to the right edge. }
				\label{fig:all_energies}
			\end{figure}

A topological phase transition is characterized by an abrupt change in a topological invariant. For the SSH model, the winding number introduced in appendix~\ref{app:winding} serves as such a topological invariant. In position space, such a topological phase transition might be continuous. Consider pure  phase B ($N=100$, $\delta = -0.15$, $a=2.0$). We may continuously transform the system from phase B to phase A by moving the left-most atom (original position
$x_1=-99.15$) to the right $x_1 = 100.85$, as sketched in Fig.~\ref{fig:sketch}.   The eigenenergies are calculated for many configurations with  $x_1 \in [-99.15,100.85]$ and plotted
 vs $x_1$  in Fig.~\ref{fig:all_energies}. The hopping elements are obtained by equation (\ref{hopping_gen}).

			One observes that mainly the lowest, the highest, and the two edge-state energies are affected. The groundstate energy is minimum if the moving atom is on top of another atom (with exceptions at the edges though). For pure phase B, the edge-state energies are almost
zero. In phase A, there are no edge-states, so that during the transition the degeneracy of the edge states is lifted. One edge state joins the valence band from above, the other the conduction band from below. It might be surprising that the degeneracy is removed only at an $x_1$-value quite close to the final phase-A position while the inversion symmetry is broken already for small shifts away from the pure phase-B configuration. 

\begin{figure} \centering
				\includegraphics[width=\columnwidth]{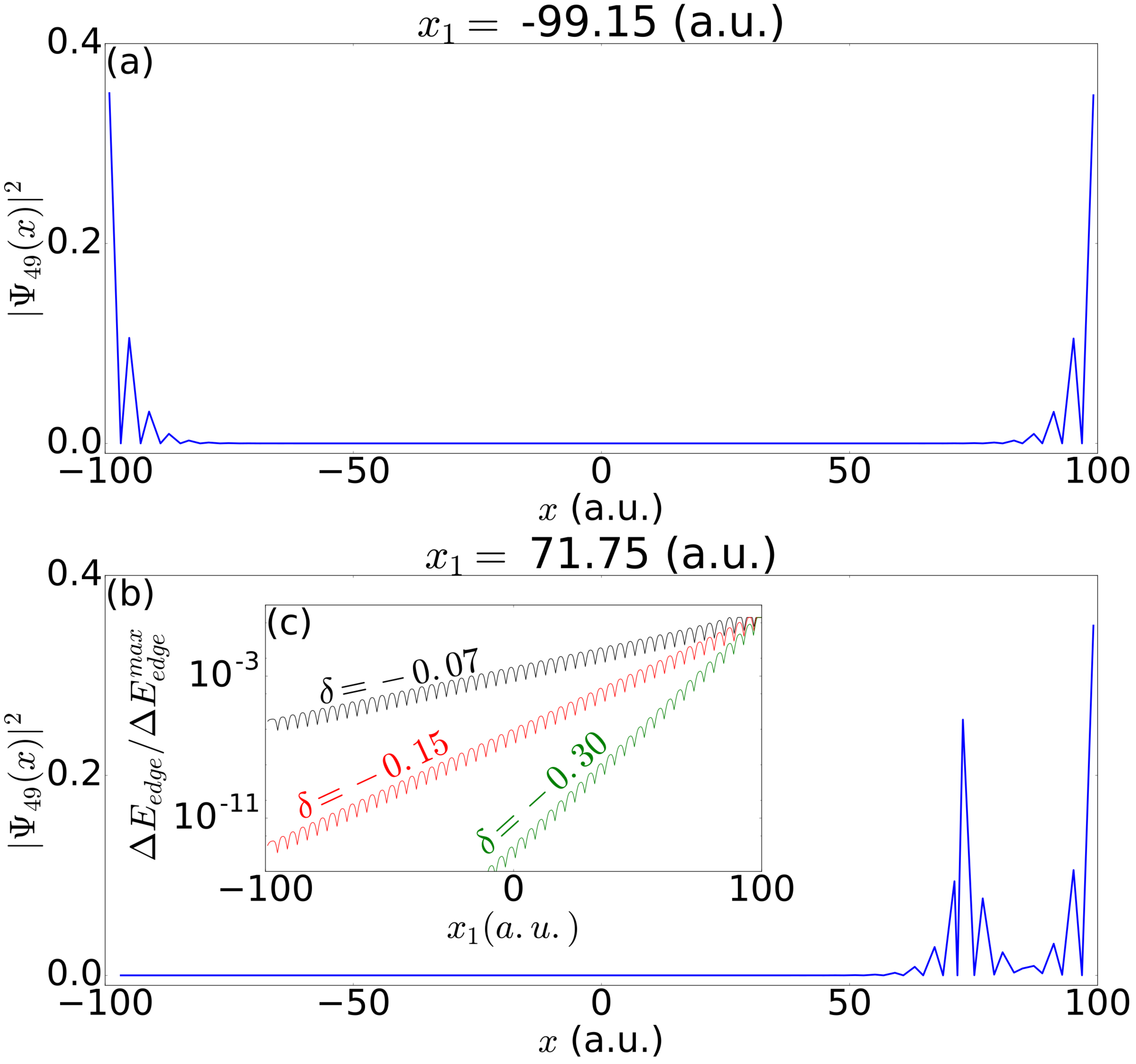}
				\caption{Probability density for edge
state $\Psi_{49}$ for different positions $x_1$. The inset shows the energy
difference between the edge states (for different $\delta$) as a function of
$x_1$.}
				\label{fig:wfct_move}
			\end{figure}

In phase B, electrons occupying the edge states are
localized at the edges, as seen in Fig.~\ref{fig:wfct_move}(a).
The left-edge part of the wavefunction moves with the moving atom
whereas the right-edge part stays at the right edge.  Plotting the energy
difference between the edge states vs $x_1$ logarithmically [see Fig.~\ref{fig:wfct_move}(c)], reveals an exponential increase. As the left part of the wavefunction  moves
towards the right side, the overlap with the right-edge part increases exponentially,  which leads to the observed exponential increase of the energy
difference. A local maximum is observed whenever the moving atom is located
at the position of another atom on sub-lattice site $\alpha =
2$. There, the wavefunction parts from the left and right interfere destructively so that for $x_1$ close to the final position at the right edge the entire wavefunction becomes delocalized.
			
		So far, we considered only two particular dimerization shifts $\delta=\pm 0.15$.	 With increasing $\delta$, the two hopping elements $v$ and $w$ differ more, and
the edge states in phase B become  more and more  localized at the edges with a smaller and smaller energy
difference. As a result, the slope of the energy difference with increasing $x_1$ is larger [see Fig.~\ref{fig:wfct_move}(c)].
			
\begin{figure} \centering
					\includegraphics[width=\columnwidth]{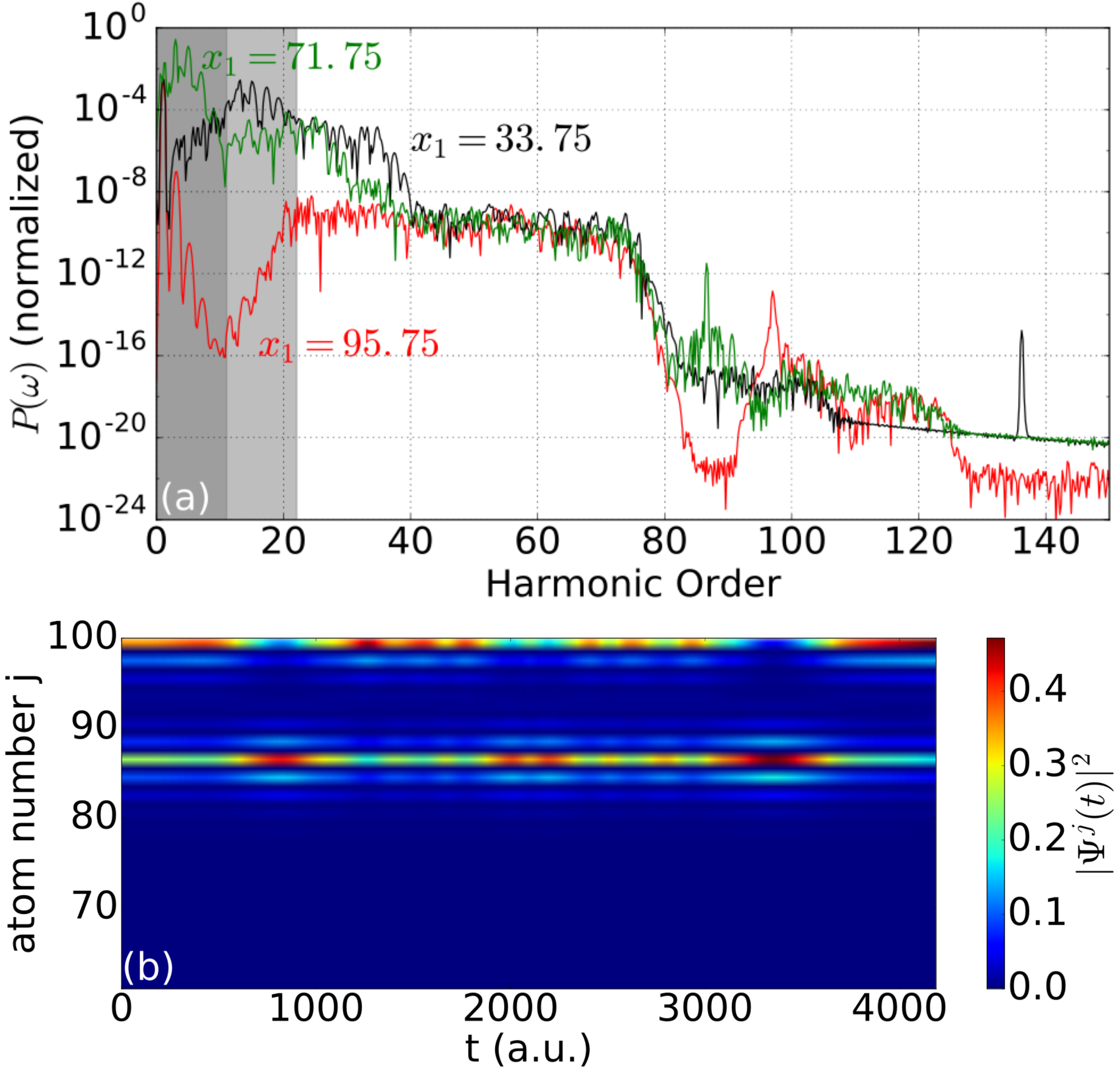}
				
				\caption{(a) Harmonic spectra for different
positions of the shifted atom $x_1$. (b) The time evolution  of
the probability density of the occupied edge state during the laser pulse for $x_1 = 71.75$.}
				\label{fig:move_spec}
			\end{figure}

As seen above, harmonic spectra for pure phase A have a strong dip for
photon energies below the band gap while those from phase B have a weaker dip for energies
below the difference between edge state energy and the bands.
The weak dip is observed up to about $x_1
\simeq 35$, as seen in Fig.~\ref{fig:move_spec}(a) for $x_1 =
33.75$. The dip is ``filled up'' for larger $x_1$ due to an increased yield for low
harmonic photon energies generated by transitions between the edge states (see
spectrum for $x_1 = 71.75$). If one neglects the edge states in the calculation of the harmonic spectrum, the dip is still there.    
The time evolution of the initially occupied edge-state orbital in the laser field is presented in Fig.~\ref{fig:move_spec}(b) and
shows a charge transfer between the right edge and the position of the shifted atom due to transitions to the other (initially not populated) edge state and
back \footnote{This is similar to transitions between almost degenerate $\sigma$-gerade and $\sigma$-ungerade states in a very much stretched diatomic molecule}.
			
                        The former edge states become delocalized once
                        their energies are shifted close to the
                        bands. Then the overall spectrum shows already the phase-A-like
                        strong dip in the sub-band-gap region, as seen
                        in Fig.~\ref{fig:move_spec}(a) for $x_1 =
                        95.75$.  Neglecting the contribution of the
                        former occupied edge state yields  a spectrum
                        without the dip for those energies because of incomplete destructive
                        interference of all the dipoles.
			
			Due to the separation of the ground state $\Psi_0$
(and the highest state $\Psi_{99}$) from the valence band (conduction band) (see Fig.~\ref{fig:all_energies}) additional
features appear in the spectrum beyond the cut-off.

		\subsection{Non-vanishing on-site potential} \label{sec:nonvanishingonsite}
An on-site potential leads to diagonal elements in the Hamiltonian matrix. 
If all diagonal elements are set to the same value $\epsilon$, the eigenfunctions $\Psi_i$ remain the same, and all eigenenergies  $E_i$ are shifted by $\epsilon$. Harmonic spectra remain unaffected by such a trivial shift of the energy scale.			
			If the diagonal elements are normally
distributed random numbers, the bandstructure is smeared out. We found that up
to a variance of $\sigma \simeq 0.05$, the dip in phase A can still
be observed.
			
\begin{figure}
	\includegraphics[width=1\columnwidth]{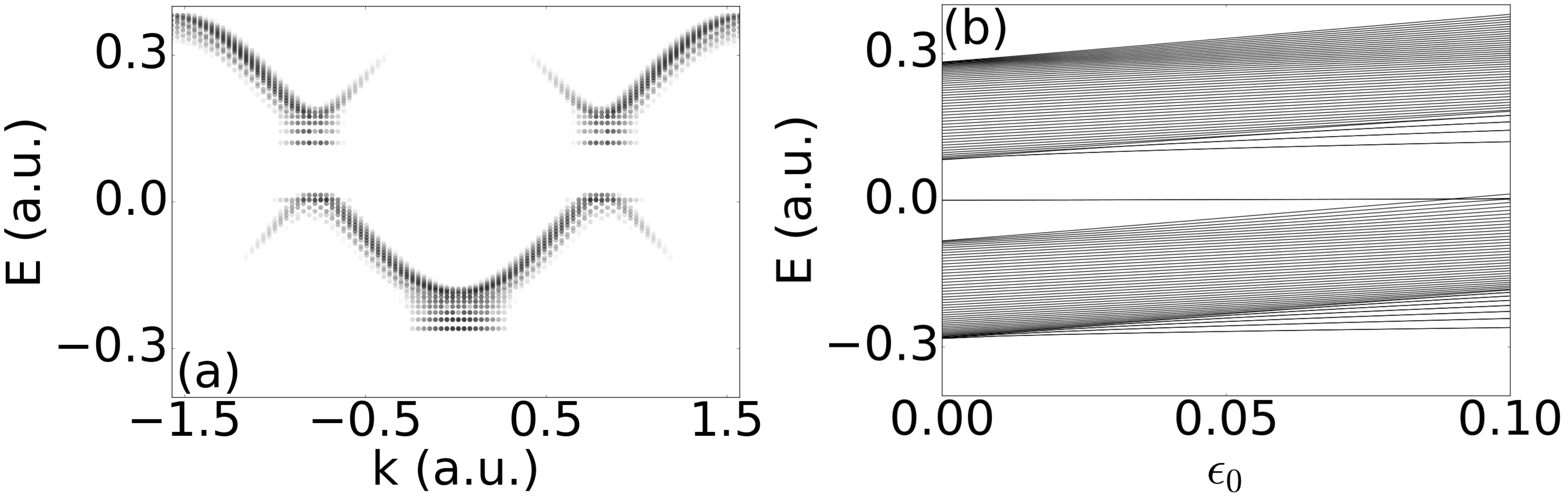}
	\caption{(a) Band structure for phase B with diagonal
		elements of frequency $\nu = 1/2$ and $\epsilon_0=0.1$. (b)  Eigenenergies  as a function of $\epsilon_0$.}
	\label{fig:Diag_B1/2}
\end{figure}

			More  interesting is a sine-shaped profile for the diagonal elements $\epsilon_j$,	
			\begin{equation} \epsilon_{j} =
\epsilon_0 \sin (2\pi\nu j/N),
			\end{equation}
which might be viewed as a generalization of the Rice-Mele model \cite{RiceMele82}.
For, e.g., $\nu= 1/2$ and $\epsilon_0=0.1$, the mean value of the $\epsilon_j$ is larger than zero so that the eigenenergies are
shifted to higher values. These shifts depend on the unperturbed energy
of the state, lower-energy states in each band are shifted less than
higher-energy states,  which leads to a separation of states from the bottoms of the
bands (see Fig.~\ref{fig:Diag_B1/2}). The chiral symmetry is broken. The separated states are
pairwise degenerate.  The
original edge states in phase B remain close to $E=0$ and thus enter the valence band around $\epsilon_0=0.09$.  For phase A, a dip in the harmonic spectrum is still
observable because of the presence of a band gap (not shown).

\begin{figure}
				\includegraphics[width=1\columnwidth]{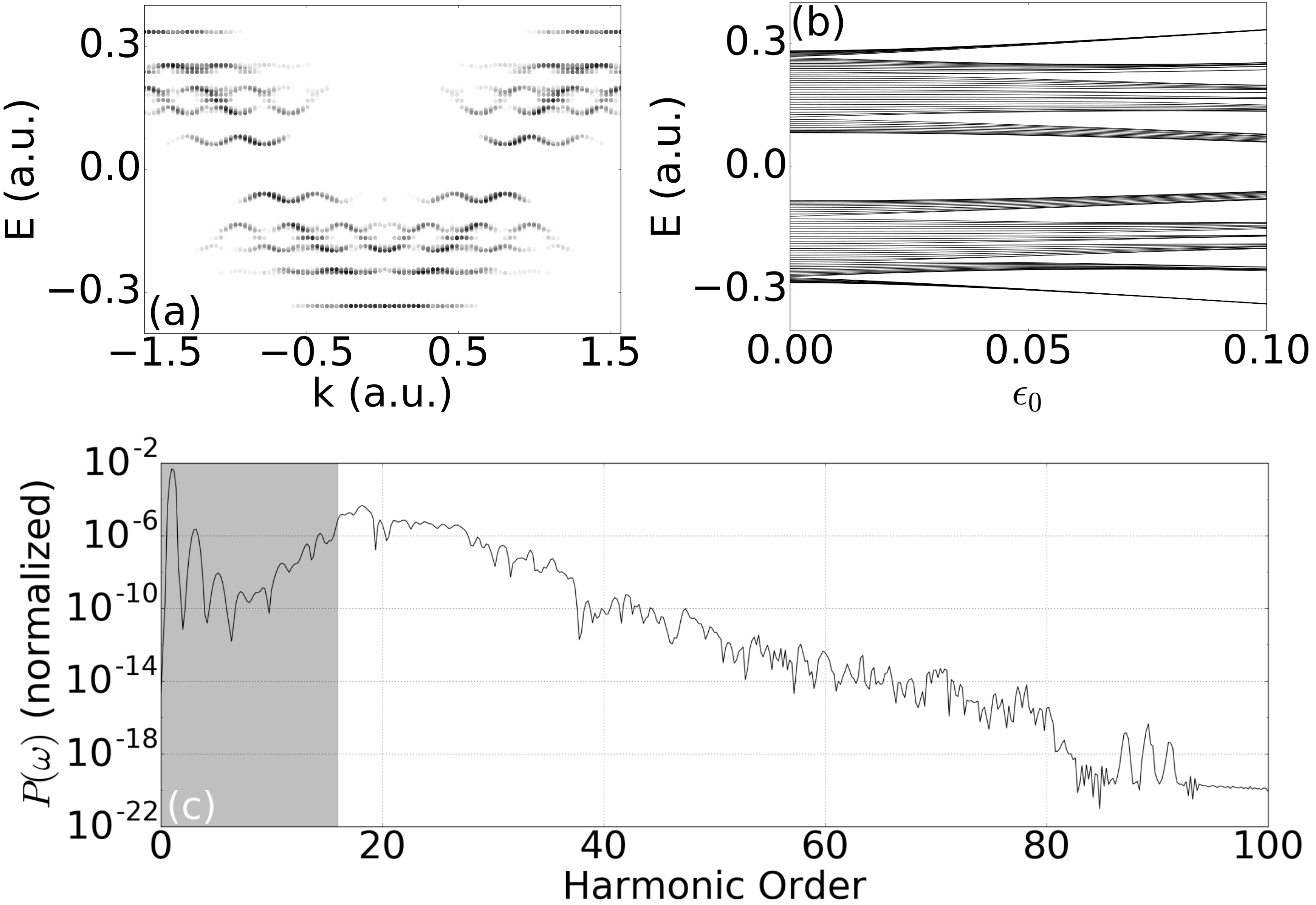}
				\caption{(a) Band structure for phase A with diagonal
elements of frequency $\nu = 10.5$ and $\epsilon_0=0.1$. (b)  Eigenenergies  as a function of $\epsilon_0$. (c) High-harmonic spectrum for $\nu = 10.5$ and $\epsilon_0=0.1$.}
				\label{fig:Diag}
			\end{figure}

For more oscillations $\nu$ in the diagonal elements, a new
periodicity is enforced on the system that increases the lattice
constant (i.e., decreases the Brillouin zone) and increases the number
of atoms per primitive cell.  For phase A, the bandstructure and the
high-harmonic spectrum is shown in Fig.~\ref{fig:Diag} for $\nu=10.5$ as well as the
evolution of the energies with increasing $\epsilon_0$. One observes a
separation of the two bands into subbands. The chiral symmetry of the energy spectrum about $E=0$ is broken [compare, for example, the second lowest and the second highest band
in Fig.~\ref{fig:Diag}(a,b)]. Nevertheless the
characteristic dip for harmonic spectra from phase A is observed because a band gap between highest occupied orbital and lowest unoccupied still exists.

	\subsection{``Measurable'' bulk-boundary correspondence?}\label{sec:BBC}
		Up to now, a laser pulse with the same intensity over the
whole chain was applied. Now the laser is focused at certain
areas of the chain according to
		\begin{equation} E(x,t) = E(t) 
\cos^2\left(\frac{(x-x_0)\pi}{2 ~x_l}\right)
		\end{equation}
		for $|x-x_0|<x_l$
		and zero otherwise. Here, $E(t)=-\partial_t A(t)$ is the previously used pulse shape in time. Note that such a tight focussing is impossible in practice because  the wavelength $\lambda \simeq 6.1
~\mathrm{\mu m}$ of the laser is large compared to the size of the chain ($9.9~\mathrm{nm}$). However, we are interested in a gedankenexperiment related to the bulk-boundary correspondence \cite{topinsRevModPhys.82.3045,PhysRevB.97.115143}. It is known that topological invariants (i.e., in our case the winding number introduced in appendix~\ref{app:winding}) are a bulk property while the presence of edge states in the band structure requires actual boundaries. Recently, it has been demonstrated for the Haldane model that the topological invariant (the Chern number) is imprinted in the phases of harmonics emitted from the bulk \cite{silva_all_2018}. However, the harmonic feature of interest in our work is the dip or its absence in the sub-band-gap harmonics for phase A and B, respectively. All our explanations relied on the presence of edge states in the band structure so that we do not expect a difference for phases A and B if the edges are not illuminated by the laser. In other words, our observable is not sensitive to the winding number but to the edge-state levels in the band structure, as is demonstrated in the following.  
		
\begin{figure} \centering
			\includegraphics[width=1\columnwidth]{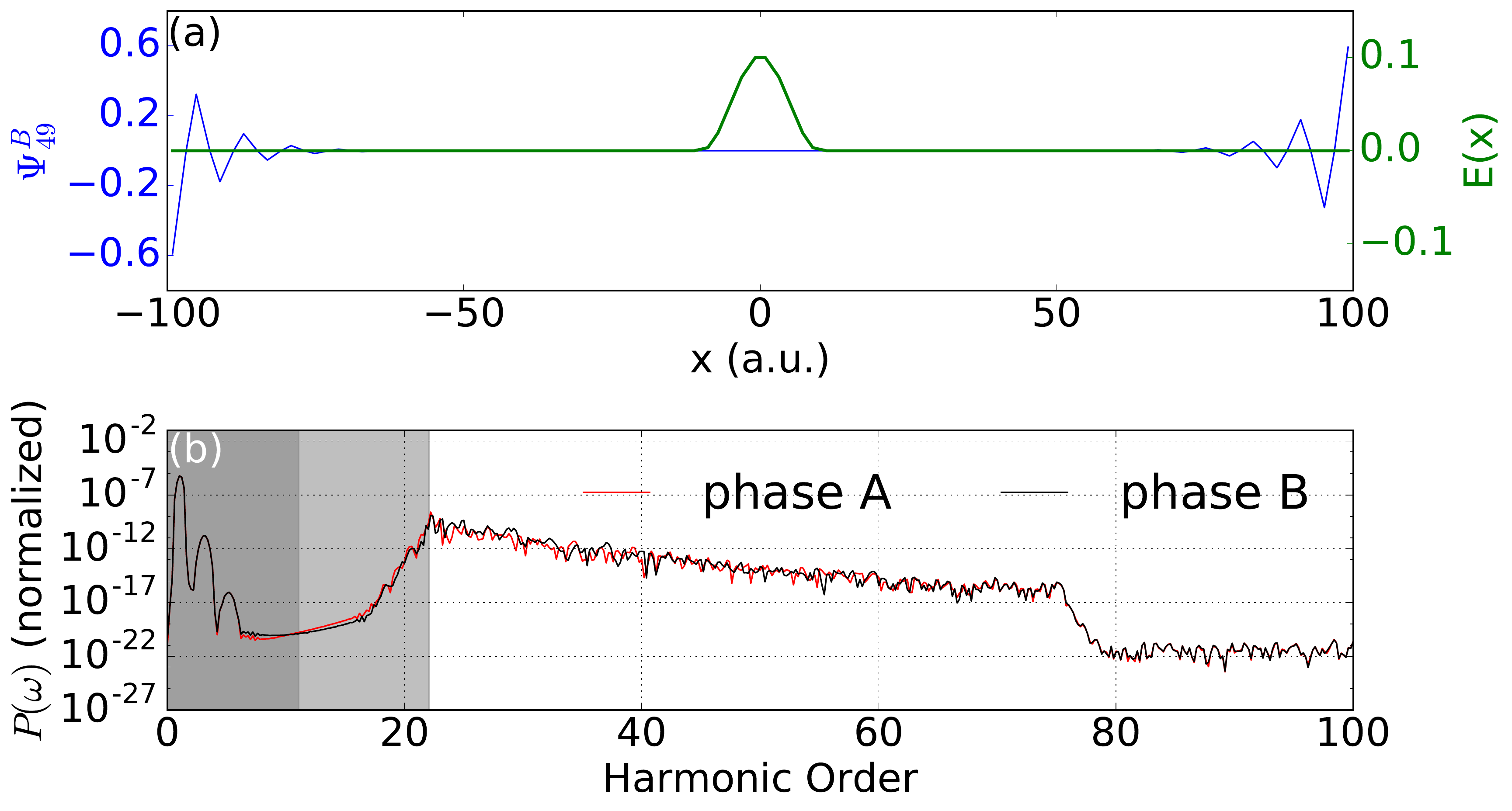}
			\caption{(a) Occupied edge-state
of phase B and the artificially narrow focus of the laser pulse in the center of the chain.   (b) High-harmonic spectra for phase A and B with the localized laser field in
the center of the chain.}
			\label{fig:local_laser}
		\end{figure}

		Using length gauge, the diagonal elements are
replaced by		
		\begin{equation} \epsilon_j=\int_{0}^{x_j}E(x,t)\, dx.
		\end{equation}	
		First, the laser pulse is focused on the center of the chain $x_0 = 0$, illuminating 10 atoms ($x_l = 10$), as indicated in Fig.~\ref{fig:local_laser}(a). The harmonic spectra for the two phases are almost identical and shown in Fig.~\ref{fig:local_laser}(b).  Especially the strong dip in the sub-band-gap region as the key feature of phase A is now also observed in phase B.
		The dip is caused by destructive interference of the individual dipoles of all
electrons in the valence band. The edge states in the band gap present in phase B  are responsible for the fact that the dip is much weaker in phase B
and at approximately half the energy. A laser illuminating only the center of the chain does not cause transitions to the edges states so
that the dip is then as strong as in phase A.
		
		\begin{figure} \centering
			\includegraphics[width=1\columnwidth]{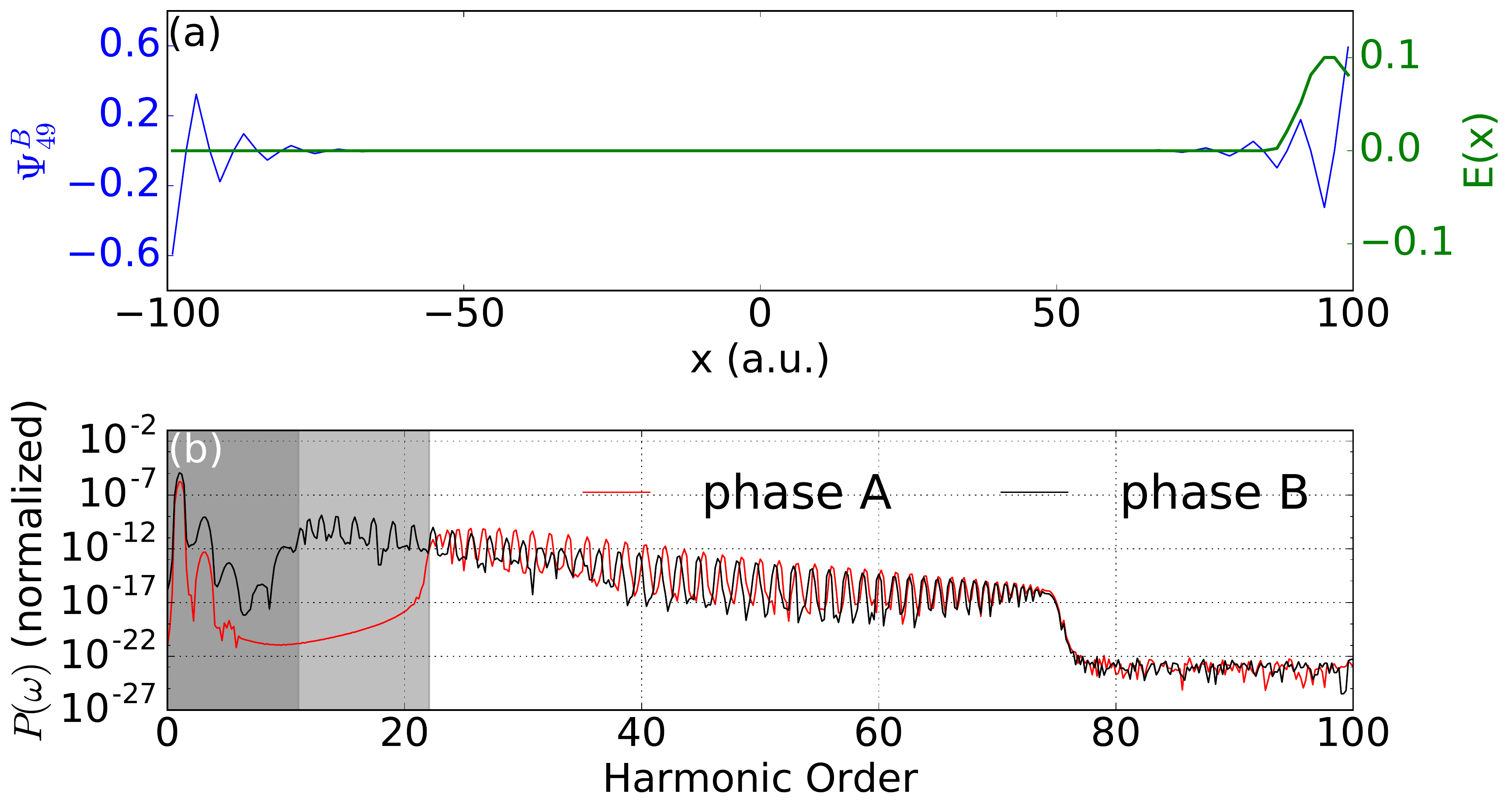}
			\caption{(a) Occupied edge-state
of phase B and the artificially narrow focus of the laser pulse on the right edge of the chain. (b)  High-harmonic spectra for phase A and B with the localized laser field at the right edge of the chain.}
			\label{fig:local_laser_edge}
		\end{figure}
		As the focus of the laser is moved towards the edges, the harmonic spectra of phase A and B become different. With the focus on, e.g., the right edge, transitions between edge states and other states become possible, which leads to harmonic spectra that are qualitatively similar to the uniformly illuminated chains, see Fig.~\ref{fig:local_laser_edge}.

	\section{Summary}
	
        High-harmonic generation in the different topological phases
        of Su-Schrieffer-Heeger chains was studied. The laser
        frequency was small compared to the band gap. In this regime,
        the overall features in the harmonic spectra can be explained
        using the common three-step model for the emission of
        harmonics above the band gap. Below-band-gap harmonics are
        strongly suppressed, causing dips in the harmonic
        spectra. Because of edge states in the middle of the
        band gap for the topological phase B, the dip is narrower, at lower harmonic orders, and less pronounced compared to phase A. As a result, a many-order-of-magnitude difference in the sub-band-gap harmonic yield between phase A and B is observed. Our results for the laser-driven SSH chain in tight-binding approximation confirm previous findings with more demanding time-dependent density functional theory simulations \cite{bauer_high-harmonic_2018}. Differences arise in the spectra for the metal phase (due to the absence of screening in the SSH modelling) and because of the absence of  high harmonics beyond the maximum energy difference between the energy levels of the SSH Hamiltonian. A remarkable robustness of the spectral features with respect to disorder in the atomic positions, a continuous transition from phase B to phase A in position space, and a modulated on-site potential was found. Further, we demonstrated that the edges need to be illuminated in order to see different harmonic spectra for phase A and phase B. However, this is not in contradiction with the bulk-boundary correspondence; rather our observable is sensitive to the presence or absence of edge states in the band structure but not to the winding number.

	\begin{appendix}
	\section{Analytical solution for the bulk Hamiltonian} \label{app:A}
	
		For the Hamiltonian of the bulk system (i.e., periodic
boundary conditions) given in equation (\ref{H_SSH_bulk}), the analytical
solution can be derived \cite{topinsshortcourse}. Rewriting the
Hamiltonian in bra-ket notation gives
				\begin{equation} \label{eq:HSSHapp}
		\begin{split} \hat{H}_0^{\mathrm{(bulk)}} =& v\sum_{m
= 1}^n\left( \ket{m,2}\bra{m,1} + \mathrm{h.c.}  \right)\\ &+w\sum_{m =
1}^n\left( \ket{m,2}\bra{m+1,1} + \mathrm{h.c.}  \right)
		\end{split}
		\end{equation}
		where $n=N/2$, and   $\ket{n+1}=\ket{1}$. We assume the hopping elements to be real-valued. The time-independent Schr\"odinger equation		
		\begin{equation} \hat{H}_0^{\mathrm{(bulk)}}
\ket{\Psi} = E\ket{\Psi}
		\end{equation}		
		can be solved by the Bloch-like ansatz
		\begin{equation}\label{ansat_Bloch}
		\begin{split} \ket{\Psi_i(k)} &= \ket{k} \otimes
\ket{u_i(k)}\\ &= \frac{1}{\sqrt{n}}\sum_{m=1}^{n}e^{imk}\ket{m}
\otimes \sum_{\alpha=1,2}g_i^\alpha(k)\ket{\alpha},
		\end{split}
		\end{equation}	
leading to		
		\begin{equation}
			\begin{split} E_{i}(k)\ket{\Psi_i(k)} &=
\frac{1}{\sqrt{n}}\sum_{m=1}^{n}e^{im k}\left[ v
g_i^1(k)\ket{m,2} \right. \\
& \quad + vg_i^2(k)\ket{m,1} +wg_i^1(k)\ket{m-1,2}  \\
& \quad \left. + w g_i^2(k)\ket{m+1,1}\right].
			\end{split}
		\end{equation}
		Multiplying by $\bra{m'}$ from the left gives
				\begin{equation}
		\begin{split} E_{i}(k)\ket{u_i(k)} &=
E_{i}(k)\left(g_i^1(k)\ket{1}+g_i^2(k)\ket{2}\right)\\ & =v
g_i^1(k)\ket{2} + v g_i^2(k)\ket{1}\\ &\quad +w
g_i^1(k)e^{ik}\ket{2} + w
g_i^2(k)e^{-ik}\ket{1},
		\end{split}
		\end{equation}
	 and in matrix representation		
		\begin{equation} E_i(k) \mathbf{u}_i(k) =
\mathbf{H}(k)\mathbf{u}_i(k),
		\end{equation}	
		with the Bloch-Hamiltonian and vector	
		\begin{equation}\label{Bulk_H} \mathbf{H}(k)
= \begin{pmatrix} 0&v+w e^{-ik}\\ v+w e^{ik}&0\\
		\end{pmatrix},
\quad  \mathbf{u}_i(k)=\begin{pmatrix} g_i^1(k)\\
g_i^2(k)\\
		\end{pmatrix},
		\end{equation}
		respectively.
		The dispersion relation for the SSH-bulk 
		\begin{equation}
		\begin{split} E_\pm (k) &= \pm\sqrt{(v+w e^{-ik})(v+w
e^{ik})}\\ &= \pm\sqrt{v^2 + w^2 + 2v w~ \mathrm{cos}~k}
		\end{split}
		\end{equation}
		follows.
		For either $v=0$ or $w=0$, the chain decomposes into
$n$ independent dimers with energy values $\pm v$ or $\pm w$,
independent of $k$ (flat bands).
				If $v$ and $w$ have the same sign, the smallest band
gap is located at the Brillouin-zone boundaries $k=\pm \pi$. Note that
for $v$ and $w$ having different signs, the shape of the bands stay
the same but they are shifted by $\pi$ along $k$.  We assume in the following that $v$ and $w$
are equally signed.
				The value for the smallest band gap is then
		
		\begin{equation}\label{bandgap} \Delta E = E_+(k=\pi)
- E_-(k=\pi)= 2\left| v - w\right|.
		\end{equation}
				For the metallic case $v = w$, the band gap disappears.		
Normalized eigenvectors are
				\begin{equation} \mathbf{u}_\pm(k) = \left(g_\pm^1(k),
g_\pm^2(k)\right)^\top = \frac{1}{\sqrt{2}}\left(1,\frac{E_\pm(k)}{v+ w
e^{-ik}}\right)^\top.
		\end{equation}
				The lowest and highest energies are		
		\begin{align} E_{\min} = E_-(0) = -\left| v - w\right|,
		\end{align}
		\begin{align} E_{\max} = E_+(0) = \left| v - w\right|,
		\end{align}	
		respectively. Insertion into the ansatz
(\ref{ansat_Bloch}) yields the corresponding states
		\begin{equation} \ket{\Psi_\pm(0)} =
\frac{1}{\sqrt{2n}}\sum_{m=1}^{n}\left(\ket{m,1}\pm\mathrm{sgn}(v+w)\ket{m,2}\right).
		\end{equation}

		\section{Winding number} \label{app:winding}
				The $2\times 2$ Bloch-Hamiltonian (\ref{Bulk_H}) can be written in the form \cite{topinsshortcourse}
		\begin{equation}\label{Ansatz_win} \mathbf{H}(k) = \mathbf{d}(k) \cdot \boldsymbol{\sigma},
		\end{equation}
				where $\boldsymbol{\sigma} = (\boldsymbol{\sigma}_x,\boldsymbol{\sigma}_y,\boldsymbol{\sigma}_z)^\top$ is the vector of  Pauli-matrices and
$\mathbf{d}(k) = \left(d_x(k),d_y(k),d_y(k)\right)^\top$ is a
3-dimensional vector, parametrized by $k$.  By comparing real and
imaginary parts of the Hamiltonian in (\ref{Bulk_H}) and
(\ref{Ansatz_win}) one finds
		\begin{equation} \mathbf{d}(k) =
\left(v+w~\mathrm{cos}(k), w~\mathrm{sin}(k),0\right)^\top,
		\end{equation}
	describing a circle of radius $|w|$ in the $d_xd_y$-plane, centered at $(v,0,0)^\top$.  
The winding number is defined as the number of times the origin is encircled  counter-clockwise as $k$ goes, e.g., from $-\pi$ to $\pi$. 
Depending on the ratio of $v$ to $w$ and their signs, the winding number can be either
$-1$ (a single clockwise encircling of the origin), $0$ (no encircling), $+1$ (a single counter-clockwise encircling of the origin), or ill-defined (in the metallic case $v=w$). For the SSH-model, the winding number is a topological
invariant because a non-vanishing winding number for the bulk ensures the presence of edge states in the finite system  (bulk-boundary correspondence) \cite{PhysRevB.97.115143}.

\bigskip

		\section{Explicit check of the gauge invariance} \label{app:gauge}
		Starting point is  the time-dependent Schr\"odinger equation (TDSE) in velocity gauge		
		\begin{equation} i \frac{\partial}{\partial
t} \mathbf{\Psi}(t) = \mathbf{H_\mathrm{\bf VG}}(t)
\mathbf{\Psi}(t). \label{eq:TDSE}
		\end{equation}
		Multiplying from the left by
		the unitary operator
	\begin{equation} \mathbf{ U}(t) = \exp[i A(t)
\mathbf{ x} ], \label{eq:U}
		\end{equation} where $\mathbf{x} = \mathrm{diag}\{x_1,x_2,
\ldots x_N\}$, gives
		\begin{equation} i \mathbf{ U}(t)
\frac{\partial}{\partial t} \mathbf{\Psi}(t) = \mathbf{ U}(t) \mathbf{H_\mathrm{\bf VG}}(t)
\mathbf{ U}^\dagger(t) \mathbf{U}(t) \mathbf{\Psi}(t). \label{eq:TDSE2}
		\end{equation} Introducing
		\begin{equation} \mathbf{\Psi}'(t) = \mathbf{U}(t) \mathbf{\Psi}(t),
		\end{equation}
			we obtain
                with \eqref{eq:U} 
			and \eqref{E_A}		
		
		\begin{equation} i \frac{\partial}{\partial
t} \mathbf{\Psi}'(t) = \left[ \mathbf{U}(t) \mathbf{H_\mathrm{VG}}(t) \mathbf{U}^\dagger(t)
+ E(t) \mathbf{x} \right]\mathbf{\Psi}'(t) .\label{eq:TDSE6}
		\end{equation}
		
		Since (suppressing the time argument and restricting
ourselves to $N=4$ for illustration)
				\begin{widetext}
			\begin{align*} \mathbf{U} \mathbf{H_\mathrm{\bf VG}} \mathbf{U}^\dagger &= 
\left( \begin{array}{ccccc} 0 & v e^{i (a-2\delta+x_1-x_2) A}
& & & \\ v e^{-i (a-2\delta+x_1-x_2) A} & 0 & w e^{i
(a+2\delta+x_2-x_3) A} & & \\ & w e^{-i (a+2\delta+x_2-x_3)
A} & 0 & v e^{i (a-2\delta+x_3-x_4) A} \\ & & v
e^{-i (a-2\delta+x_3-x_4) A} & 0
			\end{array} \right)
			\end{align*}
		\end{widetext} we see that because of
		\begin{equation} x_j-x_{j+1} =
\left\{ \begin{array}{ccc} 2\delta-a & \mathrm{if} & j \ \
\mathrm{odd} \\ -2\delta-a & \mathrm{if} & j \ \
\mathrm{even} \end{array}\right.
		\end{equation}
		simply
		\begin{equation} \mathbf{U}(t) \mathbf{H_\mathrm{\bf VG}}(t) \mathbf{U}^\dagger(t) = \mathbf{H_0} \end{equation} results, and thus
				\begin{equation} i \frac{\partial}{\partial
t} \mathbf{\Psi}'(t) = \left[ \mathbf{ H_0} + E(t) \mathbf{ x}
\right]\mathbf{\Psi}'(t) \label{eq:TDSE6.5}
		\end{equation}
follows, i.e., the length-gauge TDSE		
		\begin{equation} i \frac{\partial}{\partial
t} \mathbf{\Psi}'(t) = \mathbf{H_\mathrm{\bf LG}}(t) \mathbf{\Psi}'(t) .\label{eq:TDSE7}
		\end{equation}

\end{appendix}

	\bibliography{biblio.bib}

\begin{thebibliography}{59}%
\makeatletter
\providecommand \@ifxundefined [1]{%
 \@ifx{#1\undefined}
}%
\providecommand \@ifnum [1]{%
 \ifnum #1\expandafter \@firstoftwo
 \else \expandafter \@secondoftwo
 \fi
}%
\providecommand \@ifx [1]{%
 \ifx #1\expandafter \@firstoftwo
 \else \expandafter \@secondoftwo
 \fi
}%
\providecommand \natexlab [1]{#1}%
\providecommand \enquote  [1]{``#1''}%
\providecommand \bibnamefont  [1]{#1}%
\providecommand \bibfnamefont [1]{#1}%
\providecommand \citenamefont [1]{#1}%
\providecommand \href@noop [0]{\@secondoftwo}%
\providecommand \href [0]{\begingroup \@sanitize@url \@href}%
\providecommand \@href[1]{\@@startlink{#1}\@@href}%
\providecommand \@@href[1]{\endgroup#1\@@endlink}%
\providecommand \@sanitize@url [0]{\catcode `\\12\catcode `\$12\catcode
  `\&12\catcode `\#12\catcode `\^12\catcode `\_12\catcode `\%12\relax}%
\providecommand \@@startlink[1]{}%
\providecommand \@@endlink[0]{}%
\providecommand \url  [0]{\begingroup\@sanitize@url \@url }%
\providecommand \@url [1]{\endgroup\@href {#1}{\urlprefix }}%
\providecommand \urlprefix  [0]{URL }%
\providecommand \Eprint [0]{\href }%
\providecommand \doibase [0]{http://dx.doi.org/}%
\providecommand \selectlanguage [0]{\@gobble}%
\providecommand \bibinfo  [0]{\@secondoftwo}%
\providecommand \bibfield  [0]{\@secondoftwo}%
\providecommand \translation [1]{[#1]}%
\providecommand \BibitemOpen [0]{}%
\providecommand \bibitemStop [0]{}%
\providecommand \bibitemNoStop [0]{.\EOS\space}%
\providecommand \EOS [0]{\spacefactor3000\relax}%
\providecommand \BibitemShut  [1]{\csname bibitem#1\endcsname}%
\let\auto@bib@innerbib\@empty
\bibitem [{\citenamefont {Ghimire}\ \emph {et~al.}(2011)\citenamefont
  {Ghimire}, \citenamefont {DiChiara}, \citenamefont {Sistrunk}, \citenamefont
  {Agostini}, \citenamefont {DiMauro},\ and\ \citenamefont
  {Reis}}]{Ghimire2011}%
  \BibitemOpen
  \bibfield  {author} {\bibinfo {author} {\bibfnamefont {Shambhu}\ \bibnamefont
  {Ghimire}}, \bibinfo {author} {\bibfnamefont {Anthony~D.}\ \bibnamefont
  {DiChiara}}, \bibinfo {author} {\bibfnamefont {Emily}\ \bibnamefont
  {Sistrunk}}, \bibinfo {author} {\bibfnamefont {Pierre}\ \bibnamefont
  {Agostini}}, \bibinfo {author} {\bibfnamefont {Louis~F.}\ \bibnamefont
  {DiMauro}}, \ and\ \bibinfo {author} {\bibfnamefont {David~A.}\ \bibnamefont
  {Reis}},\ }\bibfield  {title} {\enquote {\bibinfo {title} {Observation of
  high-order harmonic generation in a bulk crystal},}\ }\href {\doibase
  10.1038/nphys1847} {\bibfield  {journal} {\bibinfo  {journal} {Nat Phys}\
  }\textbf {\bibinfo {volume} {7}},\ \bibinfo {pages} {138--141} (\bibinfo
  {year} {2011})}\BibitemShut {NoStop}%
\bibitem [{\citenamefont {Schubert}\ \emph {et~al.}(2014)\citenamefont
  {Schubert}, \citenamefont {Hohenleutner}, \citenamefont {Langer},
  \citenamefont {Urbanek}, \citenamefont {Lange}, \citenamefont {Huttner},
  \citenamefont {Golde}, \citenamefont {Meier}, \citenamefont {Kira},
  \citenamefont {Koch},\ and\ \citenamefont {Huber}}]{SchubertO.2014}%
  \BibitemOpen
  \bibfield  {author} {\bibinfo {author} {\bibfnamefont {O.}~\bibnamefont
  {Schubert}}, \bibinfo {author} {\bibfnamefont {M.}~\bibnamefont
  {Hohenleutner}}, \bibinfo {author} {\bibfnamefont {F.}~\bibnamefont
  {Langer}}, \bibinfo {author} {\bibfnamefont {B.}~\bibnamefont {Urbanek}},
  \bibinfo {author} {\bibfnamefont {C.}~\bibnamefont {Lange}}, \bibinfo
  {author} {\bibfnamefont {U.}~\bibnamefont {Huttner}}, \bibinfo {author}
  {\bibfnamefont {D.}~\bibnamefont {Golde}}, \bibinfo {author} {\bibfnamefont
  {T.}~\bibnamefont {Meier}}, \bibinfo {author} {\bibfnamefont
  {M.}~\bibnamefont {Kira}}, \bibinfo {author} {\bibfnamefont {S.W.}\
  \bibnamefont {Koch}}, \ and\ \bibinfo {author} {\bibfnamefont
  {R.}~\bibnamefont {Huber}},\ }\bibfield  {title} {\enquote {\bibinfo {title}
  {Sub-cycle control of terahertz high-harmonic generation by dynamical {Bloch}
  oscillations},}\ }\href {http://dx.doi.org/10.1038/nphoton.2013.349}
  {\bibfield  {journal} {\bibinfo  {journal} {Nat Photon}\ }\textbf {\bibinfo
  {volume} {8}},\ \bibinfo {pages} {119--123} (\bibinfo {year}
  {2014})}\BibitemShut {NoStop}%
\bibitem [{\citenamefont {Vampa}\ \emph {et~al.}(2015)\citenamefont {Vampa},
  \citenamefont {Hammond}, \citenamefont {Thir\'e}, \citenamefont {Schmidt},
  \citenamefont {L\'egar\'e}, \citenamefont {McDonald}, \citenamefont {Brabec},
  \citenamefont {Klug},\ and\ \citenamefont
  {Corkum}}]{VampaPhysRevLett.115.193603}%
  \BibitemOpen
  \bibfield  {author} {\bibinfo {author} {\bibfnamefont {G.}~\bibnamefont
  {Vampa}}, \bibinfo {author} {\bibfnamefont {T.~J.}\ \bibnamefont {Hammond}},
  \bibinfo {author} {\bibfnamefont {N.}~\bibnamefont {Thir\'e}}, \bibinfo
  {author} {\bibfnamefont {B.~E.}\ \bibnamefont {Schmidt}}, \bibinfo {author}
  {\bibfnamefont {F.}~\bibnamefont {L\'egar\'e}}, \bibinfo {author}
  {\bibfnamefont {C.~R.}\ \bibnamefont {McDonald}}, \bibinfo {author}
  {\bibfnamefont {T.}~\bibnamefont {Brabec}}, \bibinfo {author} {\bibfnamefont
  {D.~D.}\ \bibnamefont {Klug}}, \ and\ \bibinfo {author} {\bibfnamefont
  {P.~B.}\ \bibnamefont {Corkum}},\ }\bibfield  {title} {\enquote {\bibinfo
  {title} {All-optical reconstruction of crystal band structure},}\ }\href
  {\doibase 10.1103/PhysRevLett.115.193603} {\bibfield  {journal} {\bibinfo
  {journal} {Phys. Rev. Lett.}\ }\textbf {\bibinfo {volume} {115}},\ \bibinfo
  {pages} {193603} (\bibinfo {year} {2015})}\BibitemShut {NoStop}%
\bibitem [{\citenamefont {Hohenleutner}\ \emph {et~al.}(2015)\citenamefont
  {Hohenleutner}, \citenamefont {Langer}, \citenamefont {Schubert},
  \citenamefont {Knorr}, \citenamefont {Huttner}, \citenamefont {Koch},
  \citenamefont {Kira},\ and\ \citenamefont {Huber}}]{Hohenleutner2015}%
  \BibitemOpen
  \bibfield  {author} {\bibinfo {author} {\bibfnamefont {M.}~\bibnamefont
  {Hohenleutner}}, \bibinfo {author} {\bibfnamefont {F.}~\bibnamefont
  {Langer}}, \bibinfo {author} {\bibfnamefont {O.}~\bibnamefont {Schubert}},
  \bibinfo {author} {\bibfnamefont {M.}~\bibnamefont {Knorr}}, \bibinfo
  {author} {\bibfnamefont {U.}~\bibnamefont {Huttner}}, \bibinfo {author}
  {\bibfnamefont {S.~W.}\ \bibnamefont {Koch}}, \bibinfo {author}
  {\bibfnamefont {M.}~\bibnamefont {Kira}}, \ and\ \bibinfo {author}
  {\bibfnamefont {R.}~\bibnamefont {Huber}},\ }\bibfield  {title} {\enquote
  {\bibinfo {title} {Real-time observation of interfering crystal electrons in
  high-harmonic generation},}\ }\href {http://dx.doi.org/10.1038/nature14652}
  {\bibfield  {journal} {\bibinfo  {journal} {Nature}\ }\textbf {\bibinfo
  {volume} {523}},\ \bibinfo {pages} {572--575} (\bibinfo {year}
  {2015})}\BibitemShut {NoStop}%
\bibitem [{\citenamefont {Luu}\ \emph {et~al.}(2015)\citenamefont {Luu},
  \citenamefont {Garg}, \citenamefont {Kruchinin}, \citenamefont {Moulet},
  \citenamefont {Hassan},\ and\ \citenamefont {Goulielmakis}}]{Luu2015}%
  \BibitemOpen
  \bibfield  {author} {\bibinfo {author} {\bibfnamefont {T.~T.}\ \bibnamefont
  {Luu}}, \bibinfo {author} {\bibfnamefont {M.}~\bibnamefont {Garg}}, \bibinfo
  {author} {\bibfnamefont {S.~Yu}\ \bibnamefont {Kruchinin}}, \bibinfo {author}
  {\bibfnamefont {A.}~\bibnamefont {Moulet}}, \bibinfo {author} {\bibfnamefont
  {M.~Th}\ \bibnamefont {Hassan}}, \ and\ \bibinfo {author} {\bibfnamefont
  {E.}~\bibnamefont {Goulielmakis}},\ }\bibfield  {title} {\enquote {\bibinfo
  {title} {Extreme ultraviolet high-harmonic spectroscopy of solids},}\ }\href
  {http://dx.doi.org/10.1038/nature14456} {\bibfield  {journal} {\bibinfo
  {journal} {Nature}\ }\textbf {\bibinfo {volume} {521}},\ \bibinfo {pages}
  {498--502} (\bibinfo {year} {2015})}\BibitemShut {NoStop}%
\bibitem [{\citenamefont {Ndabashimiye}\ \emph {et~al.}(2016)\citenamefont
  {Ndabashimiye}, \citenamefont {Ghimire}, \citenamefont {Wu}, \citenamefont
  {Browne}, \citenamefont {Schafer}, \citenamefont {Gaarde},\ and\
  \citenamefont {Reis}}]{ndabashimiye_solid-state_2016}%
  \BibitemOpen
  \bibfield  {author} {\bibinfo {author} {\bibfnamefont {Georges}\ \bibnamefont
  {Ndabashimiye}}, \bibinfo {author} {\bibfnamefont {Shambhu}\ \bibnamefont
  {Ghimire}}, \bibinfo {author} {\bibfnamefont {Mengxi}\ \bibnamefont {Wu}},
  \bibinfo {author} {\bibfnamefont {Dana~A.}\ \bibnamefont {Browne}}, \bibinfo
  {author} {\bibfnamefont {Kenneth~J.}\ \bibnamefont {Schafer}}, \bibinfo
  {author} {\bibfnamefont {Mette~B.}\ \bibnamefont {Gaarde}}, \ and\ \bibinfo
  {author} {\bibfnamefont {David~A.}\ \bibnamefont {Reis}},\ }\bibfield
  {title} {\enquote {\bibinfo {title} {Solid-state harmonics beyond the atomic
  limit},}\ }\href {\doibase 10.1038/nature17660} {\bibfield  {journal}
  {\bibinfo  {journal} {Nature}\ }\textbf {\bibinfo {volume} {534}},\ \bibinfo
  {pages} {520--523} (\bibinfo {year} {2016})}\BibitemShut {NoStop}%
\bibitem [{\citenamefont {Langer}\ \emph {et~al.}(2017)\citenamefont {Langer},
  \citenamefont {Hohenleutner}, \citenamefont {Huttner}, \citenamefont {Koch},
  \citenamefont {Kira},\ and\ \citenamefont {Huber}}]{LangerF.2017}%
  \BibitemOpen
  \bibfield  {author} {\bibinfo {author} {\bibfnamefont {F.}~\bibnamefont
  {Langer}}, \bibinfo {author} {\bibfnamefont {M.}~\bibnamefont
  {Hohenleutner}}, \bibinfo {author} {\bibfnamefont {U.}~\bibnamefont
  {Huttner}}, \bibinfo {author} {\bibfnamefont {S.W.}\ \bibnamefont {Koch}},
  \bibinfo {author} {\bibfnamefont {M.}~\bibnamefont {Kira}}, \ and\ \bibinfo
  {author} {\bibfnamefont {R.}~\bibnamefont {Huber}},\ }\bibfield  {title}
  {\enquote {\bibinfo {title} {Symmetry-controlled temporal structure of
  high-harmonic carrier fields from a bulk crystal},}\ }\href
  {http://dx.doi.org/10.1038/nphoton.2017.29} {\bibfield  {journal} {\bibinfo
  {journal} {Nat Photon}\ }\textbf {\bibinfo {volume} {11}},\ \bibinfo {pages}
  {227--231} (\bibinfo {year} {2017})}\BibitemShut {NoStop}%
\bibitem [{\citenamefont {Tancogne-Dejean}\ \emph {et~al.}(2017)\citenamefont
  {Tancogne-Dejean}, \citenamefont {M\"ucke}, \citenamefont {K\"artner},\ and\
  \citenamefont {Rubio}}]{TancPhysRevLett.118.087403}%
  \BibitemOpen
  \bibfield  {author} {\bibinfo {author} {\bibfnamefont {Nicolas}\ \bibnamefont
  {Tancogne-Dejean}}, \bibinfo {author} {\bibfnamefont {Oliver~D.}\
  \bibnamefont {M\"ucke}}, \bibinfo {author} {\bibfnamefont {Franz~X.}\
  \bibnamefont {K\"artner}}, \ and\ \bibinfo {author} {\bibfnamefont {Angel}\
  \bibnamefont {Rubio}},\ }\bibfield  {title} {\enquote {\bibinfo {title}
  {Impact of the electronic band structure in high-harmonic generation spectra
  of solids},}\ }\href {\doibase 10.1103/PhysRevLett.118.087403} {\bibfield
  {journal} {\bibinfo  {journal} {Phys. Rev. Lett.}\ }\textbf {\bibinfo
  {volume} {118}},\ \bibinfo {pages} {087403} (\bibinfo {year}
  {2017})}\BibitemShut {NoStop}%
\bibitem [{\citenamefont {You}\ \emph {et~al.}(2017)\citenamefont {You},
  \citenamefont {Yin}, \citenamefont {Wu}, \citenamefont {Chew}, \citenamefont
  {Ren}, \citenamefont {Zhuang}, \citenamefont {Gholam-Mirzaei}, \citenamefont
  {Chini}, \citenamefont {Chang},\ and\ \citenamefont
  {Ghimire}}]{you_high-harmonic_2017}%
  \BibitemOpen
  \bibfield  {author} {\bibinfo {author} {\bibfnamefont {Yong~Sing}\
  \bibnamefont {You}}, \bibinfo {author} {\bibfnamefont {Yanchun}\ \bibnamefont
  {Yin}}, \bibinfo {author} {\bibfnamefont {Yi}~\bibnamefont {Wu}}, \bibinfo
  {author} {\bibfnamefont {Andrew}\ \bibnamefont {Chew}}, \bibinfo {author}
  {\bibfnamefont {Xiaoming}\ \bibnamefont {Ren}}, \bibinfo {author}
  {\bibfnamefont {Fengjiang}\ \bibnamefont {Zhuang}}, \bibinfo {author}
  {\bibfnamefont {Shima}\ \bibnamefont {Gholam-Mirzaei}}, \bibinfo {author}
  {\bibfnamefont {Michael}\ \bibnamefont {Chini}}, \bibinfo {author}
  {\bibfnamefont {Zenghu}\ \bibnamefont {Chang}}, \ and\ \bibinfo {author}
  {\bibfnamefont {Shambhu}\ \bibnamefont {Ghimire}},\ }\bibfield  {title}
  {\enquote {\bibinfo {title} {High-harmonic generation in amorphous solids},}\
  }\href {\doibase 10.1038/s41467-017-00989-4} {\bibfield  {journal} {\bibinfo
  {journal} {Nature Communications}\ }\textbf {\bibinfo {volume} {8}},\
  \bibinfo {pages} {724} (\bibinfo {year} {2017})}\BibitemShut {NoStop}%
\bibitem [{\citenamefont {Zhang}\ \emph {et~al.}(2018)\citenamefont {Zhang},
  \citenamefont {Si}, \citenamefont {Murakami}, \citenamefont {Bai},\ and\
  \citenamefont {George}}]{Zhang2018}%
  \BibitemOpen
  \bibfield  {author} {\bibinfo {author} {\bibfnamefont {G.~P.}\ \bibnamefont
  {Zhang}}, \bibinfo {author} {\bibfnamefont {M.~S.}\ \bibnamefont {Si}},
  \bibinfo {author} {\bibfnamefont {M.}~\bibnamefont {Murakami}}, \bibinfo
  {author} {\bibfnamefont {Y.~H.}\ \bibnamefont {Bai}}, \ and\ \bibinfo
  {author} {\bibfnamefont {Thomas~F.}\ \bibnamefont {George}},\ }\bibfield
  {title} {\enquote {\bibinfo {title} {Generating high-order optical and spin
  harmonics from ferromagnetic monolayers},}\ }\href {\doibase
  10.1038/s41467-018-05535-4} {\bibfield  {journal} {\bibinfo  {journal}
  {Nature Communications}\ }\textbf {\bibinfo {volume} {9}},\ \bibinfo {pages}
  {3031} (\bibinfo {year} {2018})}\BibitemShut {NoStop}%
\bibitem [{\citenamefont {Vampa}\ \emph {et~al.}(2018)\citenamefont {Vampa},
  \citenamefont {Hammond}, \citenamefont {Taucer}, \citenamefont {Ding},
  \citenamefont {Ropagnol}, \citenamefont {Ozaki}, \citenamefont {Delprat},
  \citenamefont {Chaker}, \citenamefont {Thir{\'e}}, \citenamefont {Schmidt},
  \citenamefont {L{\'e}gar{\'e}}, \citenamefont {Klug}, \citenamefont {Naumov},
  \citenamefont {Villeneuve}, \citenamefont {Staudte},\ and\ \citenamefont
  {Corkum}}]{Vampa2018}%
  \BibitemOpen
  \bibfield  {author} {\bibinfo {author} {\bibfnamefont {G.}~\bibnamefont
  {Vampa}}, \bibinfo {author} {\bibfnamefont {T.~J.}\ \bibnamefont {Hammond}},
  \bibinfo {author} {\bibfnamefont {M.}~\bibnamefont {Taucer}}, \bibinfo
  {author} {\bibfnamefont {Xiaoyan}\ \bibnamefont {Ding}}, \bibinfo {author}
  {\bibfnamefont {X.}~\bibnamefont {Ropagnol}}, \bibinfo {author}
  {\bibfnamefont {T.}~\bibnamefont {Ozaki}}, \bibinfo {author} {\bibfnamefont
  {S.}~\bibnamefont {Delprat}}, \bibinfo {author} {\bibfnamefont
  {M.}~\bibnamefont {Chaker}}, \bibinfo {author} {\bibfnamefont
  {N.}~\bibnamefont {Thir{\'e}}}, \bibinfo {author} {\bibfnamefont {B.~E.}\
  \bibnamefont {Schmidt}}, \bibinfo {author} {\bibfnamefont {F.}~\bibnamefont
  {L{\'e}gar{\'e}}}, \bibinfo {author} {\bibfnamefont {D.~D.}\ \bibnamefont
  {Klug}}, \bibinfo {author} {\bibfnamefont {A.~Yu}\ \bibnamefont {Naumov}},
  \bibinfo {author} {\bibfnamefont {D.~M.}\ \bibnamefont {Villeneuve}},
  \bibinfo {author} {\bibfnamefont {A.}~\bibnamefont {Staudte}}, \ and\
  \bibinfo {author} {\bibfnamefont {P.~B.}\ \bibnamefont {Corkum}},\ }\bibfield
   {title} {\enquote {\bibinfo {title} {Strong-field optoelectronics in
  solids},}\ }\href {\doibase 10.1038/s41566-018-0193-5} {\bibfield  {journal}
  {\bibinfo  {journal} {Nature Photonics}\ }\textbf {\bibinfo {volume} {12}},\
  \bibinfo {pages} {465--468} (\bibinfo {year} {2018})}\BibitemShut {NoStop}%
\bibitem [{\citenamefont {Baudisch}\ \emph {et~al.}(2018)\citenamefont
  {Baudisch}, \citenamefont {Marini}, \citenamefont {Cox}, \citenamefont {Zhu},
  \citenamefont {Silva}, \citenamefont {Teichmann}, \citenamefont {Massicotte},
  \citenamefont {Koppens}, \citenamefont {Levitov}, \citenamefont
  {Garc{\'i}a~de Abajo},\ and\ \citenamefont {Biegert}}]{Baudisch2018}%
  \BibitemOpen
  \bibfield  {author} {\bibinfo {author} {\bibfnamefont {Matthias}\
  \bibnamefont {Baudisch}}, \bibinfo {author} {\bibfnamefont {Andrea}\
  \bibnamefont {Marini}}, \bibinfo {author} {\bibfnamefont {Joel~D.}\
  \bibnamefont {Cox}}, \bibinfo {author} {\bibfnamefont {Tony}\ \bibnamefont
  {Zhu}}, \bibinfo {author} {\bibfnamefont {Francisco}\ \bibnamefont {Silva}},
  \bibinfo {author} {\bibfnamefont {Stephan}\ \bibnamefont {Teichmann}},
  \bibinfo {author} {\bibfnamefont {Mathieu}\ \bibnamefont {Massicotte}},
  \bibinfo {author} {\bibfnamefont {Frank}\ \bibnamefont {Koppens}}, \bibinfo
  {author} {\bibfnamefont {Leonid~S.}\ \bibnamefont {Levitov}}, \bibinfo
  {author} {\bibfnamefont {F.~Javier}\ \bibnamefont {Garc{\'i}a~de Abajo}}, \
  and\ \bibinfo {author} {\bibfnamefont {Jens}\ \bibnamefont {Biegert}},\
  }\bibfield  {title} {\enquote {\bibinfo {title} {Ultrafast nonlinear optical
  response of {Dirac} fermions in graphene},}\ }\href {\doibase
  10.1038/s41467-018-03413-7} {\bibfield  {journal} {\bibinfo  {journal}
  {Nature Communications}\ }\textbf {\bibinfo {volume} {9}},\ \bibinfo {pages}
  {1018} (\bibinfo {year} {2018})}\BibitemShut {NoStop}%
\bibitem [{\citenamefont {Garg}\ \emph {et~al.}(2018)\citenamefont {Garg},
  \citenamefont {Kim},\ and\ \citenamefont {Goulielmakis}}]{Garg2018}%
  \BibitemOpen
  \bibfield  {author} {\bibinfo {author} {\bibfnamefont {M.}~\bibnamefont
  {Garg}}, \bibinfo {author} {\bibfnamefont {H.~Y.}\ \bibnamefont {Kim}}, \
  and\ \bibinfo {author} {\bibfnamefont {E.}~\bibnamefont {Goulielmakis}},\
  }\bibfield  {title} {\enquote {\bibinfo {title} {Ultimate waveform
  reproducibility of extreme-ultraviolet pulses by high-harmonic generation in
  quartz},}\ }\href {\doibase 10.1038/s41566-018-0123-6} {\bibfield  {journal}
  {\bibinfo  {journal} {Nature Photonics}\ }\textbf {\bibinfo {volume} {12}},\
  \bibinfo {pages} {291--296} (\bibinfo {year} {2018})}\BibitemShut {NoStop}%
\bibitem [{\citenamefont {Hasan}\ and\ \citenamefont
  {Kane}(2010)}]{topinsRevModPhys.82.3045}%
  \BibitemOpen
  \bibfield  {author} {\bibinfo {author} {\bibfnamefont {M.~Z.}\ \bibnamefont
  {Hasan}}\ and\ \bibinfo {author} {\bibfnamefont {C.~L.}\ \bibnamefont
  {Kane}},\ }\bibfield  {title} {\enquote {\bibinfo {title} {Colloquium:
  Topological insulators},}\ }\href {\doibase 10.1103/RevModPhys.82.3045}
  {\bibfield  {journal} {\bibinfo  {journal} {Rev. Mod. Phys.}\ }\textbf
  {\bibinfo {volume} {82}},\ \bibinfo {pages} {3045--3067} (\bibinfo {year}
  {2010})}\BibitemShut {NoStop}%
\bibitem [{\citenamefont {Franz}\ and\ \citenamefont
  {Molenkamp}(2013)}]{topins}%
  \BibitemOpen
  \bibinfo {editor} {\bibfnamefont {Marcel}\ \bibnamefont {Franz}}\ and\
  \bibinfo {editor} {\bibfnamefont {Laurens}\ \bibnamefont {Molenkamp}},\
  eds.,\ \href {\doibase https://doi.org/10.1016/B978-0-444-63314-9.00012-3}
  {\emph {\bibinfo {title} {Topological Insulators}}},\ \bibinfo {series}
  {Contemporary Concepts of Condensed Matter Science}, Vol.~\bibinfo {volume}
  {6}\ (\bibinfo  {publisher} {Elsevier},\ \bibinfo {year} {2013})\BibitemShut
  {NoStop}%
\bibitem [{\citenamefont {Asb{\'o}th}\ \emph {et~al.}(2016)\citenamefont
  {Asb{\'o}th}, \citenamefont {Oroszl{\'a}ny},\ and\ \citenamefont
  {P\'alyi}}]{topinsshortcourse}%
  \BibitemOpen
  \bibfield  {author} {\bibinfo {author} {\bibfnamefont {J.K.}\ \bibnamefont
  {Asb{\'o}th}}, \bibinfo {author} {\bibfnamefont {L.}~\bibnamefont
  {Oroszl{\'a}ny}}, \ and\ \bibinfo {author} {\bibfnamefont {A.}~\bibnamefont
  {P\'alyi}},\ }\href@noop {} {\emph {\bibinfo {title} {A Short Course on
  Topological Insulators}}},\ \bibinfo {series} {Lecture Notes in Physics},
  Vol.\ \bibinfo {volume} {919}\ (\bibinfo  {publisher} {Springer},\ \bibinfo
  {year} {2016})\BibitemShut {NoStop}%
\bibitem [{\citenamefont {Qi}\ and\ \citenamefont
  {Zhang}(2011)}]{TopolinsandsupconRevModPhys.83.1057}%
  \BibitemOpen
  \bibfield  {author} {\bibinfo {author} {\bibfnamefont {Xiao-Liang}\
  \bibnamefont {Qi}}\ and\ \bibinfo {author} {\bibfnamefont {Shou-Cheng}\
  \bibnamefont {Zhang}},\ }\bibfield  {title} {\enquote {\bibinfo {title}
  {Topological insulators and superconductors},}\ }\href {\doibase
  10.1103/RevModPhys.83.1057} {\bibfield  {journal} {\bibinfo  {journal} {Rev.
  Mod. Phys.}\ }\textbf {\bibinfo {volume} {83}},\ \bibinfo {pages}
  {1057--1110} (\bibinfo {year} {2011})}\BibitemShut {NoStop}%
\bibitem [{\citenamefont {Viyuela}\ \emph {et~al.}(2018)\citenamefont
  {Viyuela}, \citenamefont {Rivas}, \citenamefont {Gasparinetti}, \citenamefont
  {Wallraff}, \citenamefont {Filipp},\ and\ \citenamefont
  {Martin-Delgado}}]{Viyuela2018}%
  \BibitemOpen
  \bibfield  {author} {\bibinfo {author} {\bibfnamefont {O.}~\bibnamefont
  {Viyuela}}, \bibinfo {author} {\bibfnamefont {A.}~\bibnamefont {Rivas}},
  \bibinfo {author} {\bibfnamefont {S.}~\bibnamefont {Gasparinetti}}, \bibinfo
  {author} {\bibfnamefont {A.}~\bibnamefont {Wallraff}}, \bibinfo {author}
  {\bibfnamefont {S.}~\bibnamefont {Filipp}}, \ and\ \bibinfo {author}
  {\bibfnamefont {M.~A.}\ \bibnamefont {Martin-Delgado}},\ }\bibfield  {title}
  {\enquote {\bibinfo {title} {Observation of topological {Uhlmann} phases with
  superconducting qubits},}\ }\href {\doibase 10.1038/s41534-017-0056-9}
  {\bibfield  {journal} {\bibinfo  {journal} {npj Quantum Information}\
  }\textbf {\bibinfo {volume} {4}},\ \bibinfo {pages} {10} (\bibinfo {year}
  {2018})}\BibitemShut {NoStop}%
\bibitem [{\citenamefont {Atala}\ \emph {et~al.}(2013)\citenamefont {Atala},
  \citenamefont {Aidelsburger}, \citenamefont {Barreiro}, \citenamefont
  {Abanin}, \citenamefont {Kitagawa}, \citenamefont {Demler},\ and\
  \citenamefont {Bloch}}]{Atala2013}%
  \BibitemOpen
  \bibfield  {author} {\bibinfo {author} {\bibfnamefont {Marcos}\ \bibnamefont
  {Atala}}, \bibinfo {author} {\bibfnamefont {Monika}\ \bibnamefont
  {Aidelsburger}}, \bibinfo {author} {\bibfnamefont {Julio~T.}\ \bibnamefont
  {Barreiro}}, \bibinfo {author} {\bibfnamefont {Dmitry}\ \bibnamefont
  {Abanin}}, \bibinfo {author} {\bibfnamefont {Takuya}\ \bibnamefont
  {Kitagawa}}, \bibinfo {author} {\bibfnamefont {Eugene}\ \bibnamefont
  {Demler}}, \ and\ \bibinfo {author} {\bibfnamefont {Immanuel}\ \bibnamefont
  {Bloch}},\ }\bibfield  {title} {\enquote {\bibinfo {title} {Direct
  measurement of the {Zak} phase in topological {Bloch} bands},}\ }\href
  {https://doi.org/10.1038/nphys2790} {\bibfield  {journal} {\bibinfo
  {journal} {Nature Physics}\ }\textbf {\bibinfo {volume} {9}},\ \bibinfo
  {pages} {795} (\bibinfo {year} {2013})}\BibitemShut {NoStop}%
\bibitem [{\citenamefont {Rechtsman}\ \emph {et~al.}(2013)\citenamefont
  {Rechtsman}, \citenamefont {Zeuner}, \citenamefont {Plotnik}, \citenamefont
  {Lumer}, \citenamefont {Podolsky}, \citenamefont {Dreisow}, \citenamefont
  {Nolte}, \citenamefont {Segev},\ and\ \citenamefont
  {Szameit}}]{Rechtsman2013}%
  \BibitemOpen
  \bibfield  {author} {\bibinfo {author} {\bibfnamefont {Mikael~C.}\
  \bibnamefont {Rechtsman}}, \bibinfo {author} {\bibfnamefont {Julia~M.}\
  \bibnamefont {Zeuner}}, \bibinfo {author} {\bibfnamefont {Yonatan}\
  \bibnamefont {Plotnik}}, \bibinfo {author} {\bibfnamefont {Yaakov}\
  \bibnamefont {Lumer}}, \bibinfo {author} {\bibfnamefont {Daniel}\
  \bibnamefont {Podolsky}}, \bibinfo {author} {\bibfnamefont {Felix}\
  \bibnamefont {Dreisow}}, \bibinfo {author} {\bibfnamefont {Stefan}\
  \bibnamefont {Nolte}}, \bibinfo {author} {\bibfnamefont {Mordechai}\
  \bibnamefont {Segev}}, \ and\ \bibinfo {author} {\bibfnamefont {Alexander}\
  \bibnamefont {Szameit}},\ }\bibfield  {title} {\enquote {\bibinfo {title}
  {Photonic {Floquet} topological insulators},}\ }\href
  {http://dx.doi.org/10.1038/nature12066} {\bibfield  {journal} {\bibinfo
  {journal} {Nature}\ }\textbf {\bibinfo {volume} {496}},\ \bibinfo {pages}
  {196} (\bibinfo {year} {2013})}\BibitemShut {NoStop}%
\bibitem [{\citenamefont {St{\"u}tzer}\ \emph {et~al.}(2018)\citenamefont
  {St{\"u}tzer}, \citenamefont {Plotnik}, \citenamefont {Lumer}, \citenamefont
  {Titum}, \citenamefont {Lindner}, \citenamefont {Segev}, \citenamefont
  {Rechtsman},\ and\ \citenamefont {Szameit}}]{Stutzer2018}%
  \BibitemOpen
  \bibfield  {author} {\bibinfo {author} {\bibfnamefont {Simon}\ \bibnamefont
  {St{\"u}tzer}}, \bibinfo {author} {\bibfnamefont {Yonatan}\ \bibnamefont
  {Plotnik}}, \bibinfo {author} {\bibfnamefont {Yaakov}\ \bibnamefont {Lumer}},
  \bibinfo {author} {\bibfnamefont {Paraj}\ \bibnamefont {Titum}}, \bibinfo
  {author} {\bibfnamefont {Netanel~H.}\ \bibnamefont {Lindner}}, \bibinfo
  {author} {\bibfnamefont {Mordechai}\ \bibnamefont {Segev}}, \bibinfo {author}
  {\bibfnamefont {Mikael~C.}\ \bibnamefont {Rechtsman}}, \ and\ \bibinfo
  {author} {\bibfnamefont {Alexander}\ \bibnamefont {Szameit}},\ }\bibfield
  {title} {\enquote {\bibinfo {title} {Photonic topological {Anderson}
  insulators},}\ }\href {\doibase 10.1038/s41586-018-0418-2} {\bibfield
  {journal} {\bibinfo  {journal} {Nature}\ }\textbf {\bibinfo {volume} {560}},\
  \bibinfo {pages} {461--465} (\bibinfo {year} {2018})}\BibitemShut {NoStop}%
\bibitem [{\citenamefont {Kruk}\ \emph {et~al.}(2019)\citenamefont {Kruk},
  \citenamefont {Poddubny}, \citenamefont {Smirnova}, \citenamefont {Wang},
  \citenamefont {Slobozhanyuk}, \citenamefont {Shorokhov}, \citenamefont
  {Kravchenko}, \citenamefont {Luther-Davies},\ and\ \citenamefont
  {Kivshar}}]{Kruk2019}%
  \BibitemOpen
  \bibfield  {author} {\bibinfo {author} {\bibfnamefont {Sergey}\ \bibnamefont
  {Kruk}}, \bibinfo {author} {\bibfnamefont {Alexander}\ \bibnamefont
  {Poddubny}}, \bibinfo {author} {\bibfnamefont {Daria}\ \bibnamefont
  {Smirnova}}, \bibinfo {author} {\bibfnamefont {Lei}\ \bibnamefont {Wang}},
  \bibinfo {author} {\bibfnamefont {Alexey}\ \bibnamefont {Slobozhanyuk}},
  \bibinfo {author} {\bibfnamefont {Alexander}\ \bibnamefont {Shorokhov}},
  \bibinfo {author} {\bibfnamefont {Ivan}\ \bibnamefont {Kravchenko}}, \bibinfo
  {author} {\bibfnamefont {Barry}\ \bibnamefont {Luther-Davies}}, \ and\
  \bibinfo {author} {\bibfnamefont {Yuri}\ \bibnamefont {Kivshar}},\ }\bibfield
   {title} {\enquote {\bibinfo {title} {Nonlinear light generation in
  topological nanostructures},}\ }\href {\doibase 10.1038/s41565-018-0324-7}
  {\bibfield  {journal} {\bibinfo  {journal} {Nature Nanotechnology}\ }\textbf
  {\bibinfo {volume} {14}},\ \bibinfo {pages} {126--130} (\bibinfo {year}
  {2019})}\BibitemShut {NoStop}%
\bibitem [{\citenamefont {Ningyuan}\ \emph {et~al.}(2015)\citenamefont
  {Ningyuan}, \citenamefont {Owens}, \citenamefont {Sommer}, \citenamefont
  {Schuster},\ and\ \citenamefont {Simon}}]{PhysRevX.5.021031}%
  \BibitemOpen
  \bibfield  {author} {\bibinfo {author} {\bibfnamefont {Jia}\ \bibnamefont
  {Ningyuan}}, \bibinfo {author} {\bibfnamefont {Clai}\ \bibnamefont {Owens}},
  \bibinfo {author} {\bibfnamefont {Ariel}\ \bibnamefont {Sommer}}, \bibinfo
  {author} {\bibfnamefont {David}\ \bibnamefont {Schuster}}, \ and\ \bibinfo
  {author} {\bibfnamefont {Jonathan}\ \bibnamefont {Simon}},\ }\bibfield
  {title} {\enquote {\bibinfo {title} {Time- and site-resolved dynamics in a
  topological circuit},}\ }\href {\doibase 10.1103/PhysRevX.5.021031}
  {\bibfield  {journal} {\bibinfo  {journal} {Phys. Rev. X}\ }\textbf {\bibinfo
  {volume} {5}},\ \bibinfo {pages} {021031} (\bibinfo {year}
  {2015})}\BibitemShut {NoStop}%
\bibitem [{\citenamefont {Albert}\ \emph {et~al.}(2015)\citenamefont {Albert},
  \citenamefont {Glazman},\ and\ \citenamefont
  {Jiang}}]{PhysRevLett.114.173902}%
  \BibitemOpen
  \bibfield  {author} {\bibinfo {author} {\bibfnamefont {Victor~V.}\
  \bibnamefont {Albert}}, \bibinfo {author} {\bibfnamefont {Leonid~I.}\
  \bibnamefont {Glazman}}, \ and\ \bibinfo {author} {\bibfnamefont {Liang}\
  \bibnamefont {Jiang}},\ }\bibfield  {title} {\enquote {\bibinfo {title}
  {Topological properties of linear circuit lattices},}\ }\href {\doibase
  10.1103/PhysRevLett.114.173902} {\bibfield  {journal} {\bibinfo  {journal}
  {Phys. Rev. Lett.}\ }\textbf {\bibinfo {volume} {114}},\ \bibinfo {pages}
  {173902} (\bibinfo {year} {2015})}\BibitemShut {NoStop}%
\bibitem [{\citenamefont {Wang}\ \emph {et~al.}(2019)\citenamefont {Wang},
  \citenamefont {Lang}, \citenamefont {Lee}, \citenamefont {Zhang},\ and\
  \citenamefont {Chong}}]{Wang2019}%
  \BibitemOpen
  \bibfield  {author} {\bibinfo {author} {\bibfnamefont {You}\ \bibnamefont
  {Wang}}, \bibinfo {author} {\bibfnamefont {Li-Jun}\ \bibnamefont {Lang}},
  \bibinfo {author} {\bibfnamefont {Ching~Hua}\ \bibnamefont {Lee}}, \bibinfo
  {author} {\bibfnamefont {Baile}\ \bibnamefont {Zhang}}, \ and\ \bibinfo
  {author} {\bibfnamefont {Y.~D.}\ \bibnamefont {Chong}},\ }\bibfield  {title}
  {\enquote {\bibinfo {title} {Topologically enhanced harmonic generation in a
  nonlinear transmission line metamaterial},}\ }\href {\doibase
  10.1038/s41467-019-08966-9} {\bibfield  {journal} {\bibinfo  {journal}
  {Nature Communications}\ }\textbf {\bibinfo {volume} {10}},\ \bibinfo {pages}
  {1102} (\bibinfo {year} {2019})}\BibitemShut {NoStop}%
\bibitem [{\citenamefont {Kane}\ and\ \citenamefont
  {Lubensky}(2013)}]{Kane2013}%
  \BibitemOpen
  \bibfield  {author} {\bibinfo {author} {\bibfnamefont {C.~L.}\ \bibnamefont
  {Kane}}\ and\ \bibinfo {author} {\bibfnamefont {T.~C.}\ \bibnamefont
  {Lubensky}},\ }\bibfield  {title} {\enquote {\bibinfo {title} {Topological
  boundary modes in isostatic lattices},}\ }\href
  {https://doi.org/10.1038/nphys2835} {\bibfield  {journal} {\bibinfo
  {journal} {Nature Physics}\ }\textbf {\bibinfo {volume} {10}},\ \bibinfo
  {pages} {39} (\bibinfo {year} {2013})}\BibitemShut {NoStop}%
\bibitem [{\citenamefont {Koochaki~Kelardeh}\ \emph {et~al.}(2017)\citenamefont
  {Koochaki~Kelardeh}, \citenamefont {Apalkov},\ and\ \citenamefont
  {Stockman}}]{PhysRevB.96.075409}%
  \BibitemOpen
  \bibfield  {author} {\bibinfo {author} {\bibfnamefont {Hamed}\ \bibnamefont
  {Koochaki~Kelardeh}}, \bibinfo {author} {\bibfnamefont {Vadym}\ \bibnamefont
  {Apalkov}}, \ and\ \bibinfo {author} {\bibfnamefont {Mark~I.}\ \bibnamefont
  {Stockman}},\ }\bibfield  {title} {\enquote {\bibinfo {title} {Graphene
  superlattices in strong circularly polarized fields: Chirality, {Berry}
  phase, and attosecond dynamics},}\ }\href {\doibase
  10.1103/PhysRevB.96.075409} {\bibfield  {journal} {\bibinfo  {journal} {Phys.
  Rev. B}\ }\textbf {\bibinfo {volume} {96}},\ \bibinfo {pages} {075409}
  (\bibinfo {year} {2017})}\BibitemShut {NoStop}%
\bibitem [{\citenamefont {Bauer}\ and\ \citenamefont
  {Hansen}(2018)}]{bauer_high-harmonic_2018}%
  \BibitemOpen
  \bibfield  {author} {\bibinfo {author} {\bibfnamefont {Dieter}\ \bibnamefont
  {Bauer}}\ and\ \bibinfo {author} {\bibfnamefont {Kenneth~K.}\ \bibnamefont
  {Hansen}},\ }\bibfield  {title} {\enquote {\bibinfo {title} {High-harmonic
  generation in solids with and without topological edge states},}\ }\href
  {\doibase 10.1103/PhysRevLett.120.177401} {\bibfield  {journal} {\bibinfo
  {journal} {Phys. Rev. Lett.}\ }\textbf {\bibinfo {volume} {120}},\ \bibinfo
  {pages} {177401} (\bibinfo {year} {2018})}\BibitemShut {NoStop}%
\bibitem [{\citenamefont {Silva}\ \emph {et~al.}()\citenamefont {Silva},
  \citenamefont {Jim\'enez-Gal\'an}, \citenamefont {Amorim}, \citenamefont
  {Smirnova},\ and\ \citenamefont {Ivanov}}]{silva_all_2018}%
  \BibitemOpen
  \bibfield  {author} {\bibinfo {author} {\bibfnamefont {R.E.F.}\ \bibnamefont
  {Silva}}, \bibinfo {author} {\bibfnamefont {\'A}\ \bibnamefont
  {Jim\'enez-Gal\'an}}, \bibinfo {author} {\bibfnamefont {B.}~\bibnamefont
  {Amorim}}, \bibinfo {author} {\bibfnamefont {O.}~\bibnamefont {Smirnova}}, \
  and\ \bibinfo {author} {\bibfnamefont {M.}~\bibnamefont {Ivanov}},\
  }\bibfield  {title} {\enquote {\bibinfo {title} {Topological strong field
  physics on sub-laser cycle time scale},}\ }\href
  {http://arxiv.org/abs/1806.11232v2} {\bibinfo  {journal}
  {{arXiv}:1806.11232v2}\ }\BibitemShut {NoStop}%
\bibitem [{\citenamefont {Chac{\'o}n}\ \emph {et~al.}()\citenamefont
  {Chac{\'o}n}, \citenamefont {Zhu}, \citenamefont {Kelly}, \citenamefont
  {Dauphin}, \citenamefont {Pisanty}, \citenamefont {Pic{\'o}n}, \citenamefont
  {Ticknor}, \citenamefont {Ciappina}, \citenamefont {Saxena},\ and\
  \citenamefont {Lewenstein}}]{chacon_observing_2018}%
  \BibitemOpen
\bibfield  {journal} {  }\bibfield  {author} {\bibinfo {author} {\bibfnamefont
  {Alexis}\ \bibnamefont {Chac{\'o}n}}, \bibinfo {author} {\bibfnamefont {Wei}\
  \bibnamefont {Zhu}}, \bibinfo {author} {\bibfnamefont {Shane~P.}\
  \bibnamefont {Kelly}}, \bibinfo {author} {\bibfnamefont {Alexandre}\
  \bibnamefont {Dauphin}}, \bibinfo {author} {\bibfnamefont {Emilio}\
  \bibnamefont {Pisanty}}, \bibinfo {author} {\bibfnamefont {Antonio}\
  \bibnamefont {Pic{\'o}n}}, \bibinfo {author} {\bibfnamefont {Christopher}\
  \bibnamefont {Ticknor}}, \bibinfo {author} {\bibfnamefont {Marcelo~F.}\
  \bibnamefont {Ciappina}}, \bibinfo {author} {\bibfnamefont {Avadh}\
  \bibnamefont {Saxena}}, \ and\ \bibinfo {author} {\bibfnamefont {Maciej}\
  \bibnamefont {Lewenstein}},\ }\bibfield  {title} {\enquote {\bibinfo {title}
  {Observing topological phase transitions with high harmonic generation},}\
  }\href {http://arxiv.org/abs/1807.01616} {\bibinfo  {journal}
  {{arXiv}:1807.01616}\ }\BibitemShut {NoStop}%
\bibitem [{\citenamefont {{Dr{\"u}eke}}\ and\ \citenamefont
  {{Bauer}}()}]{2019arXiv190101437D}%
  \BibitemOpen
\bibfield  {journal} {  }\bibfield  {author} {\bibinfo {author} {\bibfnamefont
  {Helena}\ \bibnamefont {{Dr{\"u}eke}}}\ and\ \bibinfo {author} {\bibfnamefont
  {Dieter}\ \bibnamefont {{Bauer}}},\ }\bibfield  {title} {\enquote {\bibinfo
  {title} {{Robustness of topologically sensitive harmonic generation in
  laser-driven linear chains}},}\ }\href@noop {} {\bibinfo  {journal} {accepted
  for publication in Phys. Rev. A, arXiv:1901.01437}\ }\BibitemShut {NoStop}%
\bibitem [{\citenamefont {Ikeda}\ \emph {et~al.}(2018)\citenamefont {Ikeda},
  \citenamefont {Chinzei},\ and\ \citenamefont
  {Tsunetsugu}}]{PhysRevA.98.063426}%
  \BibitemOpen
\bibfield  {journal} {  }\bibfield  {author} {\bibinfo {author} {\bibfnamefont
  {Tatsuhiko~N.}\ \bibnamefont {Ikeda}}, \bibinfo {author} {\bibfnamefont
  {Koki}\ \bibnamefont {Chinzei}}, \ and\ \bibinfo {author} {\bibfnamefont
  {Hirokazu}\ \bibnamefont {Tsunetsugu}},\ }\bibfield  {title} {\enquote
  {\bibinfo {title} {Floquet-theoretical formulation and analysis of high-order
  harmonic generation in solids},}\ }\href {\doibase
  10.1103/PhysRevA.98.063426} {\bibfield  {journal} {\bibinfo  {journal} {Phys.
  Rev. A}\ }\textbf {\bibinfo {volume} {98}},\ \bibinfo {pages} {063426}
  (\bibinfo {year} {2018})}\BibitemShut {NoStop}%
\bibitem [{\citenamefont {Luu}\ and\ \citenamefont
  {W{\"o}rner}(2018)}]{Luu2018}%
  \BibitemOpen
  \bibfield  {author} {\bibinfo {author} {\bibfnamefont {Tran~Trung}\
  \bibnamefont {Luu}}\ and\ \bibinfo {author} {\bibfnamefont {Hans~Jakob}\
  \bibnamefont {W{\"o}rner}},\ }\bibfield  {title} {\enquote {\bibinfo {title}
  {Measurement of the {Berry} curvature of solids using high-harmonic
  spectroscopy},}\ }\href {\doibase 10.1038/s41467-018-03397-4} {\bibfield
  {journal} {\bibinfo  {journal} {Nature Communications}\ }\textbf {\bibinfo
  {volume} {9}},\ \bibinfo {pages} {916} (\bibinfo {year} {2018})}\BibitemShut
  {NoStop}%
\bibitem [{\citenamefont {Reimann}\ \emph {et~al.}(2018)\citenamefont
  {Reimann}, \citenamefont {Schlauderer}, \citenamefont {Schmid}, \citenamefont
  {Langer}, \citenamefont {Baierl}, \citenamefont {Kokh}, \citenamefont
  {Tereshchenko}, \citenamefont {Kimura}, \citenamefont {Lange}, \citenamefont
  {G{\"u}dde}, \citenamefont {H{\"o}fer},\ and\ \citenamefont
  {Huber}}]{Reimann2018}%
  \BibitemOpen
  \bibfield  {author} {\bibinfo {author} {\bibfnamefont {J.}~\bibnamefont
  {Reimann}}, \bibinfo {author} {\bibfnamefont {S.}~\bibnamefont
  {Schlauderer}}, \bibinfo {author} {\bibfnamefont {C.~P.}\ \bibnamefont
  {Schmid}}, \bibinfo {author} {\bibfnamefont {F.}~\bibnamefont {Langer}},
  \bibinfo {author} {\bibfnamefont {S.}~\bibnamefont {Baierl}}, \bibinfo
  {author} {\bibfnamefont {K.~A.}\ \bibnamefont {Kokh}}, \bibinfo {author}
  {\bibfnamefont {O.~E.}\ \bibnamefont {Tereshchenko}}, \bibinfo {author}
  {\bibfnamefont {A.}~\bibnamefont {Kimura}}, \bibinfo {author} {\bibfnamefont
  {C.}~\bibnamefont {Lange}}, \bibinfo {author} {\bibfnamefont
  {J.}~\bibnamefont {G{\"u}dde}}, \bibinfo {author} {\bibfnamefont
  {U.}~\bibnamefont {H{\"o}fer}}, \ and\ \bibinfo {author} {\bibfnamefont
  {R.}~\bibnamefont {Huber}},\ }\bibfield  {title} {\enquote {\bibinfo {title}
  {Subcycle observation of lightwave-driven {Dirac} currents in a topological
  surface band},}\ }\href {\doibase 10.1038/s41586-018-0544-x} {\bibfield
  {journal} {\bibinfo  {journal} {Nature}\ }\textbf {\bibinfo {volume} {562}},\
  \bibinfo {pages} {396--400} (\bibinfo {year} {2018})}\BibitemShut {NoStop}%
\bibitem [{\citenamefont {Sommer}\ \emph {et~al.}(2016)\citenamefont {Sommer},
  \citenamefont {Bothschafter}, \citenamefont {Sato}, \citenamefont {Jakubeit},
  \citenamefont {Latka}, \citenamefont {Razskazovskaya}, \citenamefont
  {Fattahi}, \citenamefont {Jobst}, \citenamefont {Schweinberger},
  \citenamefont {Shirvanyan}, \citenamefont {Yakovlev}, \citenamefont
  {Kienberger}, \citenamefont {Yabana}, \citenamefont {Karpowicz},
  \citenamefont {Schultze},\ and\ \citenamefont {Krausz}}]{Sommer2016}%
  \BibitemOpen
  \bibfield  {author} {\bibinfo {author} {\bibfnamefont {A.}~\bibnamefont
  {Sommer}}, \bibinfo {author} {\bibfnamefont {E.~M.}\ \bibnamefont
  {Bothschafter}}, \bibinfo {author} {\bibfnamefont {S.~A.}\ \bibnamefont
  {Sato}}, \bibinfo {author} {\bibfnamefont {C.}~\bibnamefont {Jakubeit}},
  \bibinfo {author} {\bibfnamefont {T.}~\bibnamefont {Latka}}, \bibinfo
  {author} {\bibfnamefont {O.}~\bibnamefont {Razskazovskaya}}, \bibinfo
  {author} {\bibfnamefont {H.}~\bibnamefont {Fattahi}}, \bibinfo {author}
  {\bibfnamefont {M.}~\bibnamefont {Jobst}}, \bibinfo {author} {\bibfnamefont
  {W.}~\bibnamefont {Schweinberger}}, \bibinfo {author} {\bibfnamefont
  {V.}~\bibnamefont {Shirvanyan}}, \bibinfo {author} {\bibfnamefont {V.~S.}\
  \bibnamefont {Yakovlev}}, \bibinfo {author} {\bibfnamefont {R.}~\bibnamefont
  {Kienberger}}, \bibinfo {author} {\bibfnamefont {K.}~\bibnamefont {Yabana}},
  \bibinfo {author} {\bibfnamefont {N.}~\bibnamefont {Karpowicz}}, \bibinfo
  {author} {\bibfnamefont {M.}~\bibnamefont {Schultze}}, \ and\ \bibinfo
  {author} {\bibfnamefont {F.}~\bibnamefont {Krausz}},\ }\bibfield  {title}
  {\enquote {\bibinfo {title} {Attosecond nonlinear polarization and
  light-matter energy transfer in solids},}\ }\href
  {http://dx.doi.org/10.1038/nature17650} {\bibfield  {journal} {\bibinfo
  {journal} {Nature}\ }\textbf {\bibinfo {volume} {534}},\ \bibinfo {pages}
  {86--90} (\bibinfo {year} {2016})}\BibitemShut {NoStop}%
\bibitem [{\citenamefont {Garg}\ \emph {et~al.}(2016)\citenamefont {Garg},
  \citenamefont {Zhan}, \citenamefont {Luu}, \citenamefont {Lakhotia},
  \citenamefont {Klostermann}, \citenamefont {Guggenmos},\ and\ \citenamefont
  {Goulielmakis}}]{Garg2016}%
  \BibitemOpen
  \bibfield  {author} {\bibinfo {author} {\bibfnamefont {M.}~\bibnamefont
  {Garg}}, \bibinfo {author} {\bibfnamefont {M.}~\bibnamefont {Zhan}}, \bibinfo
  {author} {\bibfnamefont {T.~T.}\ \bibnamefont {Luu}}, \bibinfo {author}
  {\bibfnamefont {H.}~\bibnamefont {Lakhotia}}, \bibinfo {author}
  {\bibfnamefont {T.}~\bibnamefont {Klostermann}}, \bibinfo {author}
  {\bibfnamefont {A.}~\bibnamefont {Guggenmos}}, \ and\ \bibinfo {author}
  {\bibfnamefont {E.}~\bibnamefont {Goulielmakis}},\ }\bibfield  {title}
  {\enquote {\bibinfo {title} {Multi-petahertz electronic metrology},}\ }\href
  {http://dx.doi.org/10.1038/nature19821} {\bibfield  {journal} {\bibinfo
  {journal} {Nature}\ }\textbf {\bibinfo {volume} {538}},\ \bibinfo {pages}
  {359--363} (\bibinfo {year} {2016})}\BibitemShut {NoStop}%
\bibitem [{\citenamefont {Higuchi}\ \emph {et~al.}(2017)\citenamefont
  {Higuchi}, \citenamefont {Heide}, \citenamefont {Ullmann}, \citenamefont
  {Weber},\ and\ \citenamefont {Hommelhoff}}]{Higuchi2017}%
  \BibitemOpen
  \bibfield  {author} {\bibinfo {author} {\bibfnamefont {Takuya}\ \bibnamefont
  {Higuchi}}, \bibinfo {author} {\bibfnamefont {Christian}\ \bibnamefont
  {Heide}}, \bibinfo {author} {\bibfnamefont {Konrad}\ \bibnamefont {Ullmann}},
  \bibinfo {author} {\bibfnamefont {Heiko~B.}\ \bibnamefont {Weber}}, \ and\
  \bibinfo {author} {\bibfnamefont {Peter}\ \bibnamefont {Hommelhoff}},\
  }\bibfield  {title} {\enquote {\bibinfo {title} {Light-field-driven currents
  in graphene},}\ }\href {http://dx.doi.org/10.1038/nature23900} {\bibfield
  {journal} {\bibinfo  {journal} {Nature}\ }\textbf {\bibinfo {volume} {550}},\
  \bibinfo {pages} {224} (\bibinfo {year} {2017})}\BibitemShut {NoStop}%
\bibitem [{\citenamefont {Heide}\ \emph {et~al.}(2018)\citenamefont {Heide},
  \citenamefont {Higuchi}, \citenamefont {Weber},\ and\ \citenamefont
  {Hommelhoff}}]{PhysRevLett.121.207401}%
  \BibitemOpen
  \bibfield  {author} {\bibinfo {author} {\bibfnamefont {Christian}\
  \bibnamefont {Heide}}, \bibinfo {author} {\bibfnamefont {Takuya}\
  \bibnamefont {Higuchi}}, \bibinfo {author} {\bibfnamefont {Heiko~B.}\
  \bibnamefont {Weber}}, \ and\ \bibinfo {author} {\bibfnamefont {Peter}\
  \bibnamefont {Hommelhoff}},\ }\bibfield  {title} {\enquote {\bibinfo {title}
  {Coherent electron trajectory control in graphene},}\ }\href {\doibase
  10.1103/PhysRevLett.121.207401} {\bibfield  {journal} {\bibinfo  {journal}
  {Phys. Rev. Lett.}\ }\textbf {\bibinfo {volume} {121}},\ \bibinfo {pages}
  {207401} (\bibinfo {year} {2018})}\BibitemShut {NoStop}%
\bibitem [{\citenamefont {Runge}\ and\ \citenamefont
  {Gross}(1984)}]{rungegross84}%
  \BibitemOpen
  \bibfield  {author} {\bibinfo {author} {\bibfnamefont {Erich}\ \bibnamefont
  {Runge}}\ and\ \bibinfo {author} {\bibfnamefont {E.~K.~U.}\ \bibnamefont
  {Gross}},\ }\bibfield  {title} {\enquote {\bibinfo {title}
  {Density-functional theory for time-dependent systems},}\ }\href {\doibase
  10.1103/PhysRevLett.52.997} {\bibfield  {journal} {\bibinfo  {journal} {Phys.
  Rev. Lett.}\ }\textbf {\bibinfo {volume} {52}},\ \bibinfo {pages} {997--1000}
  (\bibinfo {year} {1984})}\BibitemShut {NoStop}%
\bibitem [{\citenamefont {Ullrich}(2011)}]{ullrich_time-dependent_2011}%
  \BibitemOpen
  \bibfield  {author} {\bibinfo {author} {\bibfnamefont {Carsten~A.}\
  \bibnamefont {Ullrich}},\ }\href {\doibase
  10.1093/acprof:oso/9780199563029.001.0001} {\emph {\bibinfo {title}
  {Time-Dependent Density-Functional Theory: Concepts and Applications}}},\
  Oxford Graduate Texts\ (\bibinfo  {publisher} {Oxford University Press},\
  \bibinfo {year} {2011})\BibitemShut {NoStop}%
\bibitem [{\citenamefont {Su}\ \emph {et~al.}(1979)\citenamefont {Su},
  \citenamefont {Schrieffer},\ and\ \citenamefont
  {Heeger}}]{SSHPhysRevLett.42.1698}%
  \BibitemOpen
  \bibfield  {author} {\bibinfo {author} {\bibfnamefont {W.~P.}\ \bibnamefont
  {Su}}, \bibinfo {author} {\bibfnamefont {J.~R.}\ \bibnamefont {Schrieffer}},
  \ and\ \bibinfo {author} {\bibfnamefont {A.~J.}\ \bibnamefont {Heeger}},\
  }\bibfield  {title} {\enquote {\bibinfo {title} {Solitons in
  polyacetylene},}\ }\href {\doibase 10.1103/PhysRevLett.42.1698} {\bibfield
  {journal} {\bibinfo  {journal} {Phys. Rev. Lett.}\ }\textbf {\bibinfo
  {volume} {42}},\ \bibinfo {pages} {1698--1701} (\bibinfo {year}
  {1979})}\BibitemShut {NoStop}%
\bibitem [{\citenamefont {Streitwolf}(1985)}]{PSSB:PSSB2221270102}%
  \BibitemOpen
  \bibfield  {author} {\bibinfo {author} {\bibfnamefont {H.~W.}\ \bibnamefont
  {Streitwolf}},\ }\bibfield  {title} {\enquote {\bibinfo {title} {Physical
  properties of polyacetylene},}\ }\href {\doibase 10.1002/pssb.2221270102}
  {\bibfield  {journal} {\bibinfo  {journal} {physica status solidi (b)}\
  }\textbf {\bibinfo {volume} {127}},\ \bibinfo {pages} {11--54} (\bibinfo
  {year} {1985})}\BibitemShut {NoStop}%
\bibitem [{\citenamefont {Block}\ and\ \citenamefont
  {Streitwolf}(1996)}]{BLOCK199631}%
  \BibitemOpen
  \bibfield  {author} {\bibinfo {author} {\bibfnamefont {S.}~\bibnamefont
  {Block}}\ and\ \bibinfo {author} {\bibfnamefont {H.W.}\ \bibnamefont
  {Streitwolf}},\ }\bibfield  {title} {\enquote {\bibinfo {title} {Calculated
  photoinduced dynamics in trans-polyacetylene},}\ }\href {\doibase
  https://doi.org/10.1016/0379-6779(95)03413-E} {\bibfield  {journal} {\bibinfo
   {journal} {Synthetic Metals}\ }\textbf {\bibinfo {volume} {76}},\ \bibinfo
  {pages} {31 -- 33} (\bibinfo {year} {1996})}\BibitemShut {NoStop}%
\bibitem [{\citenamefont {Gebhard}\ \emph {et~al.}(1997)\citenamefont
  {Gebhard}, \citenamefont {Bott}, \citenamefont {Scheidler}, \citenamefont
  {Thomas},\ and\ \citenamefont {Koch}}]{gebhardPhMB}%
  \BibitemOpen
  \bibfield  {author} {\bibinfo {author} {\bibfnamefont {F.}~\bibnamefont
  {Gebhard}}, \bibinfo {author} {\bibfnamefont {K.}~\bibnamefont {Bott}},
  \bibinfo {author} {\bibfnamefont {M.}~\bibnamefont {Scheidler}}, \bibinfo
  {author} {\bibfnamefont {P.}~\bibnamefont {Thomas}}, \ and\ \bibinfo {author}
  {\bibfnamefont {S.~W.}\ \bibnamefont {Koch}},\ }\bibfield  {title} {\enquote
  {\bibinfo {title} {Optical absorption of non-interacting tight-binding
  electrons in a peierls-distorted chain at half band-filling},}\ }\href
  {\doibase 10.1080/13642819708205700} {\bibfield  {journal} {\bibinfo
  {journal} {Philosophical Magazine B}\ }\textbf {\bibinfo {volume} {75}},\
  \bibinfo {pages} {1--12} (\bibinfo {year} {1997})}\BibitemShut {NoStop}%
\bibitem [{\citenamefont {Hansen}\ \emph {et~al.}(2017)\citenamefont {Hansen},
  \citenamefont {Deffge},\ and\ \citenamefont {Bauer}}]{PhysRevA.96.053418}%
  \BibitemOpen
  \bibfield  {author} {\bibinfo {author} {\bibfnamefont {Kenneth~K.}\
  \bibnamefont {Hansen}}, \bibinfo {author} {\bibfnamefont {Tobias}\
  \bibnamefont {Deffge}}, \ and\ \bibinfo {author} {\bibfnamefont {Dieter}\
  \bibnamefont {Bauer}},\ }\bibfield  {title} {\enquote {\bibinfo {title}
  {High-order harmonic generation in solid slabs beyond the
  single-active-electron approximation},}\ }\href {\doibase
  10.1103/PhysRevA.96.053418} {\bibfield  {journal} {\bibinfo  {journal} {Phys.
  Rev. A}\ }\textbf {\bibinfo {volume} {96}},\ \bibinfo {pages} {053418}
  (\bibinfo {year} {2017})}\BibitemShut {NoStop}%
\bibitem [{\citenamefont {Hansen}\ \emph {et~al.}(2018)\citenamefont {Hansen},
  \citenamefont {Bauer},\ and\ \citenamefont {Madsen}}]{PhysRevA.97.043424}%
  \BibitemOpen
  \bibfield  {author} {\bibinfo {author} {\bibfnamefont {Kenneth~K.}\
  \bibnamefont {Hansen}}, \bibinfo {author} {\bibfnamefont {Dieter}\
  \bibnamefont {Bauer}}, \ and\ \bibinfo {author} {\bibfnamefont {Lars~Bojer}\
  \bibnamefont {Madsen}},\ }\bibfield  {title} {\enquote {\bibinfo {title}
  {Finite-system effects on high-order harmonic generation: From atoms to
  solids},}\ }\href {\doibase 10.1103/PhysRevA.97.043424} {\bibfield  {journal}
  {\bibinfo  {journal} {Phys. Rev. A}\ }\textbf {\bibinfo {volume} {97}},\
  \bibinfo {pages} {043424} (\bibinfo {year} {2018})}\BibitemShut {NoStop}%
\bibitem [{\citenamefont {Yu}\ \emph {et~al.}(2019)\citenamefont {Yu},
  \citenamefont {Hansen},\ and\ \citenamefont {Madsen}}]{PhysRevA.99.013435}%
  \BibitemOpen
  \bibfield  {author} {\bibinfo {author} {\bibfnamefont {Chuan}\ \bibnamefont
  {Yu}}, \bibinfo {author} {\bibfnamefont {Kenneth~K.}\ \bibnamefont {Hansen}},
  \ and\ \bibinfo {author} {\bibfnamefont {Lars~Bojer}\ \bibnamefont
  {Madsen}},\ }\bibfield  {title} {\enquote {\bibinfo {title} {Enhanced
  high-order harmonic generation in donor-doped band-gap materials},}\ }\href
  {\doibase 10.1103/PhysRevA.99.013435} {\bibfield  {journal} {\bibinfo
  {journal} {Phys. Rev. A}\ }\textbf {\bibinfo {volume} {99}},\ \bibinfo
  {pages} {013435} (\bibinfo {year} {2019})}\BibitemShut {NoStop}%
\bibitem [{\citenamefont {Zvyagin}(2018)}]{PhysRevB.97.144412}%
  \BibitemOpen
  \bibfield  {author} {\bibinfo {author} {\bibfnamefont {A.~A.}\ \bibnamefont
  {Zvyagin}},\ }\bibfield  {title} {\enquote {\bibinfo {title} {Topological
  edge states and impurities: Manifestation in the local static and dynamical
  characteristics of dimerized quantum chains},}\ }\href {\doibase
  10.1103/PhysRevB.97.144412} {\bibfield  {journal} {\bibinfo  {journal} {Phys.
  Rev. B}\ }\textbf {\bibinfo {volume} {97}},\ \bibinfo {pages} {144412}
  (\bibinfo {year} {2018})}\BibitemShut {NoStop}%
\bibitem [{Note1()}]{Note1}%
  \BibitemOpen
  \bibinfo {note} {The ``band structure'' for finite SSH chains is calculated
  by Fourier-transforming the eigenstates, i.e., $\Psi _i(x)\rightarrow
  \protect \mathaccentV {tilde}07E\Psi _i(k)$, and plotting $\protect \qopname
  \relax o{log}|\protect \mathaccentV {tilde}07E\Psi _i(k)|^2$ vs $k$ and the
  respective energy $E_i$ as color-coded contours. The step size for $k$ is
  $\Delta k = \protect \frac {2\pi }{N a}$. Hence possible $k$-values are $k =
  0,\Delta k, 2\Delta k,...,(N-1)\Delta k$. The first Brillouin-zone in the
  metal case is $[-\protect \frac {\pi }{a},\protect \frac {\pi }{a}]$. In
  phases A and B the spacing between the atoms is not equidistant. However, as
  long as $|\delta | \ll a$ we can use the same procedure to calculate the band
  structures shown in Fig.~\ref {fig:bands_AB} for illustration.}\BibitemShut
  {Stop}%
\bibitem [{Note2()}]{Note2}%
  \BibitemOpen
  \bibinfo {note} {Due to the lack of chiral symmetry of the Kohn-Sham
  Hamiltonian, the conduction and valence band are neither symmetric about
  energy $E=0$ nor are the edge states in phase B exactly in the middle of the
  band gap in the DFT band structure in Ref.~\cite {bauer_high-harmonic_2018}.
  This shows already that chiral symmetry and the related existence of a
  winding number \cite {topinsshortcourse} is not necessary for a 1D chain to
  display degenerate edge states.}\BibitemShut {Stop}%
\bibitem [{\citenamefont {Graf}\ and\ \citenamefont
  {Vogl}(1995)}]{PhysRevB.51.4940}%
  \BibitemOpen
  \bibfield  {author} {\bibinfo {author} {\bibfnamefont {M.}~\bibnamefont
  {Graf}}\ and\ \bibinfo {author} {\bibfnamefont {P.}~\bibnamefont {Vogl}},\
  }\bibfield  {title} {\enquote {\bibinfo {title} {Electromagnetic fields and
  dielectric response in empirical tight-binding theory},}\ }\href {\doibase
  10.1103/PhysRevB.51.4940} {\bibfield  {journal} {\bibinfo  {journal} {Phys.
  Rev. B}\ }\textbf {\bibinfo {volume} {51}},\ \bibinfo {pages} {4940--4949}
  (\bibinfo {year} {1995})}\BibitemShut {NoStop}%
\bibitem [{\citenamefont {Bandrauk}\ \emph {et~al.}(2009)\citenamefont
  {Bandrauk}, \citenamefont {Chelkowski}, \citenamefont {Diestler},
  \citenamefont {Manz},\ and\ \citenamefont {Yuan}}]{PhysRevA.79.023403}%
  \BibitemOpen
  \bibfield  {author} {\bibinfo {author} {\bibfnamefont {A.~D.}\ \bibnamefont
  {Bandrauk}}, \bibinfo {author} {\bibfnamefont {S.}~\bibnamefont
  {Chelkowski}}, \bibinfo {author} {\bibfnamefont {D.~J.}\ \bibnamefont
  {Diestler}}, \bibinfo {author} {\bibfnamefont {J.}~\bibnamefont {Manz}}, \
  and\ \bibinfo {author} {\bibfnamefont {K.-J.}\ \bibnamefont {Yuan}},\
  }\bibfield  {title} {\enquote {\bibinfo {title} {Quantum simulation of
  high-order harmonic spectra of the hydrogen atom},}\ }\href {\doibase
  10.1103/PhysRevA.79.023403} {\bibfield  {journal} {\bibinfo  {journal} {Phys.
  Rev. A}\ }\textbf {\bibinfo {volume} {79}},\ \bibinfo {pages} {023403}
  (\bibinfo {year} {2009})}\BibitemShut {NoStop}%
\bibitem [{\citenamefont {Baggesen}\ and\ \citenamefont
  {Madsen}(2011)}]{0953-4075-44-11-115601}%
  \BibitemOpen
  \bibfield  {author} {\bibinfo {author} {\bibfnamefont {Jan~Conrad}\
  \bibnamefont {Baggesen}}\ and\ \bibinfo {author} {\bibfnamefont {Lars~Bojer}\
  \bibnamefont {Madsen}},\ }\bibfield  {title} {\enquote {\bibinfo {title} {On
  the dipole, velocity and acceleration forms in high-order harmonic generation
  from a single atom or molecule},}\ }\href
  {http://stacks.iop.org/0953-4075/44/i=11/a=115601} {\bibfield  {journal}
  {\bibinfo  {journal} {Journal of Physics B: Atomic, Molecular and Optical
  Physics}\ }\textbf {\bibinfo {volume} {44}},\ \bibinfo {pages} {115601}
  (\bibinfo {year} {2011})}\BibitemShut {NoStop}%
\bibitem [{\citenamefont {Sundaram}\ and\ \citenamefont
  {Milonni}(1990)}]{PhysRevA.41.6571}%
  \BibitemOpen
  \bibfield  {author} {\bibinfo {author} {\bibfnamefont {Bala}\ \bibnamefont
  {Sundaram}}\ and\ \bibinfo {author} {\bibfnamefont {Peter~W.}\ \bibnamefont
  {Milonni}},\ }\bibfield  {title} {\enquote {\bibinfo {title} {High-order
  harmonic generation: Simplified model and relevance of single-atom theories
  to experiment},}\ }\href {\doibase 10.1103/PhysRevA.41.6571} {\bibfield
  {journal} {\bibinfo  {journal} {Phys. Rev. A}\ }\textbf {\bibinfo {volume}
  {41}},\ \bibinfo {pages} {6571--6573} (\bibinfo {year} {1990})}\BibitemShut
  {NoStop}%
\bibitem [{\citenamefont {Rhim}\ \emph {et~al.}(2018)\citenamefont {Rhim},
  \citenamefont {Bardarson},\ and\ \citenamefont
  {Slager}}]{PhysRevB.97.115143}%
  \BibitemOpen
  \bibfield  {author} {\bibinfo {author} {\bibfnamefont {Jun-Won}\ \bibnamefont
  {Rhim}}, \bibinfo {author} {\bibfnamefont {Jens~H.}\ \bibnamefont
  {Bardarson}}, \ and\ \bibinfo {author} {\bibfnamefont {Robert-Jan}\
  \bibnamefont {Slager}},\ }\bibfield  {title} {\enquote {\bibinfo {title}
  {Unified bulk-boundary correspondence for band insulators},}\ }\href
  {\doibase 10.1103/PhysRevB.97.115143} {\bibfield  {journal} {\bibinfo
  {journal} {Phys. Rev. B}\ }\textbf {\bibinfo {volume} {97}},\ \bibinfo
  {pages} {115143} (\bibinfo {year} {2018})}\BibitemShut {NoStop}%
\bibitem [{\citenamefont {Vampa}\ and\ \citenamefont
  {Brabec}(2017)}]{vampa_merge_2017}%
  \BibitemOpen
  \bibfield  {author} {\bibinfo {author} {\bibfnamefont {G}~\bibnamefont
  {Vampa}}\ and\ \bibinfo {author} {\bibfnamefont {T}~\bibnamefont {Brabec}},\
  }\bibfield  {title} {\enquote {\bibinfo {title} {Merge of high harmonic
  generation from gases and solids and its implications for attosecond
  science},}\ }\href {\doibase 10.1088/1361-6455/aa528d} {\bibfield  {journal}
  {\bibinfo  {journal} {Journal of Physics B: Atomic, Molecular and Optical
  Physics}\ }\textbf {\bibinfo {volume} {50}},\ \bibinfo {pages} {083001}
  (\bibinfo {year} {2017})}\BibitemShut {NoStop}%
\bibitem [{Note3()}]{Note3}%
  \BibitemOpen
  \bibinfo {note} {Whereas a TDDFT simulation with ``frozen'' (i.e.,
  ground-state) Kohn-Sham potential does.}\BibitemShut {Stop}%
\bibitem [{Note4()}]{Note4}%
  \BibitemOpen
  \bibinfo {note} {This is similar to transitions between almost degenerate
  $\sigma $-gerade and $\sigma $-ungerade states in a very much stretched
  diatomic molecule}\BibitemShut {NoStop}%
\bibitem [{\citenamefont {Rice}\ and\ \citenamefont {Mele}(1982)}]{RiceMele82}%
  \BibitemOpen
  \bibfield  {author} {\bibinfo {author} {\bibfnamefont {M.~J.}\ \bibnamefont
  {Rice}}\ and\ \bibinfo {author} {\bibfnamefont {E.~J.}\ \bibnamefont
  {Mele}},\ }\bibfield  {title} {\enquote {\bibinfo {title} {Elementary
  excitations of a linearly conjugated diatomic polymer},}\ }\href {\doibase
  10.1103/PhysRevLett.49.1455} {\bibfield  {journal} {\bibinfo  {journal}
  {Phys. Rev. Lett.}\ }\textbf {\bibinfo {volume} {49}},\ \bibinfo {pages}
  {1455--1459} (\bibinfo {year} {1982})}\BibitemShut {NoStop}%
\end{thebibliography}%

\end{document}